%% file: paper.tex
\documentclass[extra]{gji}
\pdfoutput=1

\usepackage[dvipsinames]{xcolor}

\usepackage[%
            colorlinks=True,%
            urlcolor=blue,%
            linkcolor=blue,%
            citecolor= blue,%
            pdfauthor={T. Gastine, J. Aubert, A. Fournier},%
            pdftitle={Dynamo-based limit to the extent of a stable layer atop 
Earth's core},]{hyperref}
\usepackage{timet}
\usepackage{lineno}
\usepackage{xcolor}
\usepackage{amsmath,amsfonts,amssymb}
\usepackage{graphicx}
\usepackage{journals}
\usepackage{booktabs}
\usepackage{longtable}
\usepackage{caption}

\urldef\ggg\url{%
https://agupubs.onlinelibrary.wiley.com/action/downloadSupplement?doi=10.1002%2F
2016GC006438&file=ggge21074-sup-0002-2016GC006438-s02.zip}

\title{Dynamo-based limit to the extent of a stable layer atop Earth's core}

\author[T. Gastine, J. Aubert and A. Fournier]
  {Thomas Gastine\thanks{Email: 
\href{mailto:gastine@ipgp.fr}{gastine@ipgp.fr}}, Julien Aubert, Alexandre 
Fournier \\
  Institut de Physique du Globe de Paris, Sorbonne Paris Cit\'e,
Universit\'e Paris-Diderot, UMR 7154 CNRS, 1 rue Jussieu, F-75005 Paris,
France}

\date{Received \today; in original form \today}
\pagerange{\pageref{firstpage}--\pageref{lastpage}}
\volume{200}
\pubyear{2019}

\def\vec#1{\ensuremath{\mathchoice{\mbox{\boldmath$\displaystyle#1$}}
{\mbox{\boldmath$\textstyle#1$}}
{\mbox{\boldmath$\scriptstyle#1$}}
{\mbox{\boldmath$\scriptscriptstyle#1$}}}}

\newcommand{\dpen}{\mathcal{D}_p}
\newcommand{\hstrat}{\mathcal{H}_s}
\newcommand{\heff}{\mathcal{H}_\text{eff}}
\newcommand{\ecmb}{\mathcal{M}_\ell(r_o)}
\newcommand{\ask}{\alpha_{\text{SK}}}
\newcommand{\qsk}{\mathcal{Q}_\ell^\text{SK}}

\begin{document}

\label{firstpage}

\maketitle

\begin{summary}

The existence of a stably stratified layer underneath the core-mantle 
boundary (CMB) has been recently revived by corroborating evidences 
coming from seismic studies, mineral physics and thermal evolution models.
Such a layer could find its physical origination either in 
compositional stratification due to the accumulation of light elements 
at the top or the core or in thermal stratification due to the heat flux 
becoming locally sub-adiabatic. The exact 
properties of this stably-stratified layer, namely its size $\hstrat$ and the 
degree of its stratification characterised by the Brunt-V\"ais\"al\"a frequency 
$N$, are however uncertain and highly debated.
A stable layer underneath the CMB can have crucial dynamical impacts 
on the geodynamo. Because of the inhibition of the convective motions, a 
stable layer is expected to primarily act as a low-pass filter on the magnetic 
field, smoothing out the rapidly-varying and small-scale features by skin 
effect. 
To investigate this effect more systematically, we compute 70 global 
geodynamo models varying the size of the stably-stratified layer from 0 to 
300~km and its amplitude from $N/\Omega = 0$ to $N/\Omega \simeq 50$, $\Omega$ 
being the rotation rate.
We show that the penetration of the convective flow in the stably-stratified 
layer is controlled by the typical size of the 
convective eddies and by the local variations of the ratio $N/\Omega$. Using 
quantitative measures of the degree of morphological semblance between the 
magnetic field obtained 
in numerical models and the geomagnetic field at the CMB, we establish an upper 
bound for the stable layer thickness $\hstrat < (N/\Omega)^{-1} 
\mathcal{L}_s$, $\mathcal{L}_s$ being the horizontal size of the convective 
flow at the base of the stable layer. This defines a strong geomagnetic 
constraint on the properties of a stably-stratified layer beneath the CMB. 
Unless unaccounted double-diffusive effects could drastically modify the 
dynamics of the stable layer, our numerical geodynamo models hence
favour no stable stratification atop of the core.
\end{summary}


\section{Introduction}

The convective motions that develop in Earth's liquid outer core are considered 
as the primary source of power to sustain the geomagnetic field via 
dynamo action. This results from the combination of thermal and compositional 
buoyancy sources. The Earth secular cooling and the latent heat
release due to the solidification of iron at the inner core boundary (ICB) 
provide the thermal heat sources, while the expulsion of light elements 
from the iron-rich inner core into the fluid outer core constitutes another
source of buoyancy of compositional origin \citep[e.g.][]{Lister95}.

The exact convective state of the Earth liquid core is however uncertain. 
The usual assumption posits that the outer core is entirely convective, 
well-mixed by the turbulent convective motions. This hypothesis has been 
however questioned by seismic studies that rather
suggest the presence of inhomogeneous layers above the ICB 
\citep[e.g.][]{Souriau91} or below the core-mantle 
boundary (CMB) \citep[e.g.][]{Tanaka07,Helffrich10,Kaneshima18}.
Those layers could arise because of stable stratification of thermal or 
compositional origin. The degree of stratification can be quantified by the 
Brunt-V\"ais\"al\"a frequency expressed by
\begin{equation}
 N^2 = -\dfrac{g}{\rho} \dfrac{\partial \rho}{\partial r}-\dfrac{\rho g^2}{K_S},
\end{equation}
where $g$ is the gravity, $K_S$ the isentropic bulk modulus and $\rho$ the 
fluid density.
The possible stable layer underneath the CMB has been recently the focus of a 
large array of studies that span various scientific fields encompassing seismic 
studies, mineral physics and geomagnetic analyses \citep[for a review, 
see][]{Hirose13}.

On the seismology side, several studies, based on the analysis of travel times 
of SmKS waves, report $P$-wave velocities between 0.1\% and 1\% slower than 
PREM at the top of the core. They attribute this deviation to an inhomogeneous 
stably stratified layer which would yield a mean density 
profile that would significantly depart from the adiabat. The inferred 
thickness $\hstrat$ of 
this layer has evolved from $\hstrat\sim 100$~km in earlier studies 
\citep[e.g.][]{Lay90,Tanaka07} to larger values ranging from $300$ to $450$~km 
in more recent analyses \citep{Helffrich10,Tang15,Kaneshima15,Kaneshima18}.
The evaluation of the associated Brunt-V\"ais\"al\"a frequency is always 
delicate since it directly depends on the chemical composition of the core
\citep[e.g.][]{Brodholt17} but tentative estimates yield $N \sim 0.5-1~$~mHz 
\citep{Helffrich10}. There is, however, no consensus on the interpretation of 
these seismic observations, and some seismic studies rather favour no 
stratification at the top of the core \citep[e.g.][]{Alexandrakis10,Irving18}. 
\cite{Irving18} for instance explain the deviations to PREM by a refined 
equation of state that yields steeper density profiles close to the CMB.

Stable stratification of thermal origin arises when the temperature gradient
becomes sub-adiabatic. This directly depends on the heat flux at the 
core-mantle boundary and on the outer core thermal conductivity. 
The latter has been the subject of intense debates over the recent years.
\textit{Ab-initio} first principle numerical calculations yield 
conductivity values ranging from $100$ to $150$~W.m$^{-1}$.K$^{-1}$ 
\citep{deKoker12,Pozzo12,Pozzo13} significantly larger than previous estimate of 
$30$~W.m$^{-1}$.K$^{-1}$  \citep{Stacey07}.  On the other hand, high-pressure 
experiments yield contradictory results: while some are supportive of the 
\textit{ab-initio} findings \citep{Gomi13,Ohta16}, others rather favour the 
lower previously-accepted conductivity value \citep{Konopkova16}. A CMB heat 
flux of roughly $Q_\text{CMB}=15$~TW would be required to accommodate a 
fully-convective core for the highest thermal conductivities. Although 
estimates of the actual heat flux at the CMB are rather uncertain 
\citep[e.g.][]{Lay08}, $Q_\text{CMB}=15$~TW certainly lies in the high range of 
commonly-accepted values. Stable thermal stratification below the CMB is hence 
the favoured scenario \citep{Pozzo12,Gomi13}, would the actual core 
conductivity lies in the current high-range estimate. 

Geomagnetic observations provide another source of constraints on the physical 
properties of a stable layer underneath the CMB, since this layer would damp 
radial motions and/or harbour waves for which gravity would act as a restoring 
force. The geomagnetic secular variation (SV) is governed for the most part by
fluid flow at the top of the core. The presence of a stably stratified
layer underneath the core-mantle boundary implies that the radial velocity 
is weaker than the horizontal components. Using arguments
based on a careful analysis of the Navier-Stokes equations under the
tangentially geostrophic and Boussinesq approximations in a stratified
layer, \cite{Jault91} showed that the corresponding flow is not
strictly toroidal, as its large-scale components can be partly poloidal. In
short, even if the radial flow is much smaller than the horizontal one, its
radial gradient can not be neglected against the horizontal divergence of
the flow for the large scales of motion. In that sense, trying to
establish that the core is stratified considering purely toroidal core
surface flow for the analysis of the SV may be overkill, especially when
one is restricted to analyse  the large scales of motion. Accordingly, 
\cite{Lesur15} found that a large-scale core surface flow permitting up- and 
down-wellings was more adapted to account for the secular variation during the 
magnetic satellite era than its strictly toroidal equivalent precluding radial 
flow underneath the core-mantle boundary. The latter hypothesis led typically to 
a 15\% increase in the root-mean-squared misfit to low-latitude satellite data
compared to the misfit obtained with the former. There are regions at the
core surface (for instance underneath the Indian Ocean), where some radial
flow is mandatory to account for the data \citep[e.g.][]{Amit14,Baerenzung16}. 
That does not mean that there is no stratified 
layer, it simply implies that SV data alone do not have a real resolving power 
on the properties of a hypothetical
stratified layer at the top of core. In fact, in this study we shall stress
that much stronger constraints are obtained by studying the morphology of
the magnetic field at the top of the core.
With regard to wave motion, \cite{Brag93} hypothesised that the decadal 
variations of the magnetic field could be related to the excitation of MAC waves 
in a stable layer with  $\hstrat=80$~km and $N\sim \Omega$, $\Omega$ being 
Earth's rotation rate. This idea was more recently revisited by \cite{Buffett14} 
who attributes the 60~yr period observed in the secular variation of the 
axisymmetric dipole to MAC waves. Best-fitting linear models yield 
$\hstrat=130-140$~km and $N=0.74-0.84\,\Omega$ \citep{Buffett16}, a degree of 
stratification much weaker than the estimates coming from seismic studies. 
In practice, the reference models are assumed to be 
spherically-symmetric and yield a function $N(r)$. 
Table~\ref{tab:strat} lists selected publications which provide 
estimates of $\hstrat$ and $N_m/\Omega$, with $N_m=\max_r N(r)$.

\begin{table}
\centering
\caption{Selected publications that propose values for the 
physical properties of the stably-stratified layer underneath the 
CMB using $\Omega=7.29\times 10^{-5}$~s$^{-1}$.
}
\begin{tabular}{llrr}
\toprule
Reference & Name & $\hstrat$ (km) & $N_m/\Omega$ \\
\midrule
% \cite{Lay90} & LZ90 & 50-100 & 43$^*$ \\ % 1-2\%
\cite{Brag93} & B93 & 80 & 2 \\
% \cite{Tanaka07} & T07 & 110-170 & $>0$\\ % 0.8\% 
\cite{Buffet10} & BS11 & 70 & 55 \\
  \cite{Helffrich10} & HK10 & 300 & 7-14.7 \\
\cite{Gubbins13} & GD13 & 100 & 20.6 \\
%   \cite{Tang15} & TZH15 & 300 & $>5$\\ % 0.25%
\cite{Buffett16} & BKH16 & 130-140 & 0.74-0.84 \\
%   \cite{Kaneshima18} & K18 & 450 & 0.35\% 8.3$^*$ \\
\cite{Irving18} & ICL18 & 0 & $\simeq 0$ \\
\bottomrule
\end{tabular}
 \label{tab:strat}
\end{table}

% Geomagnetic observations provide another source of constraints on the physical 
% properties of a stable layer underneath the CMB.
% \cite{Brag93} hypothesised that the decadal variations of the magnetic field 
% could be related to the excitation of MAC waves in a stable layer
% with  $\hstrat=80$~km and $N\sim \Omega$, $\Omega$ being Earth's rotation 
% rate. This idea was more recently revisited by \cite{Buffett14} who attributes 
% the 60~yr period observed in the secular variation of the axisymmetric dipole 
% to MAC waves. Best-fitting linear models yield $\hstrat=130-140$~km and 
% $N=0.74-0.84\,\Omega$ \citep{Buffett16}, a degree of stratification much weaker 
% than the estimates coming from seismic studies. 
% In addition, \cite{Lesur15} found that a large-scale core surface flow
% permitting up- and downwellings was more adapted to account for the geomagnetic 
% secular variation during the magnetic satellite era than its strictly toroidal
% equivalent precluding radial flow underneath the core-mantle boundary. The
% latter hypothesis led typically to a 15\% increase in the root-mean-squared
% misfit to low-latitude satellite data compared to the misfit obtained
% with the former. 
% This puts a strong constraint on the maximum extent of a stable layer.
% The analysis by \cite{Gubbins07} for instance suggests $\hstrat \leq 100$~km,
% although this upper bound is likely to directly depend on the efficiency of the 
% convective penetration.

In the present study, we aim to analyse the physical influence of a stable 
layer below the CMB by means of 3-D global geodynamo models. 
\cite{Takehiro01} analysed the propagation of thermal Rossby waves in presence 
of a stably-stratified temperature gradient in the limit of an inviscid fluid.
They showed that the distance of penetration $\dpen$ of a 
convective eddy of size $\mathcal{L}_s$ is inversely proportional to 
the ratio of the Brunt-V\"ais\"al\"a  and the rotation frequencies
% Theoretical 
% analyses by \cite{Takehiro01} suggest that, in absence of viscosity and 
% magnetic field, the distance of penetration $\dpen$ of a convective eddy of size 
% $\mathcal{L}_s$ is directly controlled by the ratio of the 
% Brunt-V\"ais\"al\"a  and the 
\begin{equation}
 \dpen \sim \left(\dfrac{N}{\Omega}\right)^{-1} \mathcal{L}_s\,.
 \label{eq:dpen_takehiro}
\end{equation}
Hence, the larger the ratio $N/\Omega$, the smaller the penetration distance.
The above theoretical scaling can be seen as the result of two competing linear 
physical effects: on the one-hand rapid-rotation goes along wih quasi 
bi-dimensional Taylor columns aligned with the rotation axis, while on the other 
hand the stable stratification promotes motions in horizontal planes 
perpendicular to the radial stratification. 
Subsequent analyses by \cite{Takehiro15} have however questioned the 
validity of this hydrodynamical scaling relation in presence of a magnetic 
field.
Based on the penetration distance of 
Alfv\'en waves, he instead suggests that the above hydrodynamical scaling 
could be replaced by
\begin{equation}
 \dfrac{\dpen}{d} \sim \dfrac{\omega_{A}}{\omega_{\text{diss}}}\,,
\label{eq:takalu}
\end{equation}
where $\omega_A$ is the typical frequency of the Alfv\'en waves, 
$\omega_{\text{diss}}$ is a diffusion frequency resulting from the average 
between kinematic and magnetic diffusivities, and $d$ is the extent of 
the fluid domain. However, the
validity of the above linear scaling has only been tested by \cite{Takehiro18a} 
in the context of nonlinear models of rotating convection in presence of an 
imposed background magnetic field. Global 3-D numerical 
simulations of stellar \citep{Brun17} and planetary \citep{Dietrich18} 
convection in spherical shells under the anelastic approximation have shown 
little support for the hydrodynamical scaling (\ref{eq:dpen_takehiro}). This is 
likely because of the important role played by inertia in these numerical 
computations where rotation has a moderate influence on the convective flow 
\citep[e.g.][]{Zahn91,Hurlburt94}. Geodynamo models
that incoporate a stable layer are either limited to moderate degrees of 
stratification $N/\Omega< 5$ \citep{Olson17,Yan18,Christensen18} or to 
weakly supercritical convection \citep{Nakagawa15}, hence restricting further 
tests of the relevance of the above scalings. The first goal of the present 
study is precisely to estimate the penetration distance in rapidly-rotating 
geodynamo models to assess the validity of 
Eqs.~(\ref{eq:dpen_takehiro}-\ref{eq:takalu}).

Numerical dynamo models have also shown that stable layers can have a strong
impact on the magnetic field. In the limit of vanishing penetrative convection, 
a stably-stratified region can be roughly approximated by a stagnant 
conducting fluid layer. The magnetic field parts which vary rapidly with time 
are then strongly damped by the magnetic skin effect. In the context of 
modelling Mercury's dynamo, \cite{Christensen06} has for instance shown that 
the magnetic field atop a stable layer becomes more dipolar and more 
axisymmetric 
\citep[see also][]{Gubbins07,Christensen08,Stanley08,Takahashi19}. The second 
objective of this study consists in quantifying the influence of a stable layer 
on the magnetic field morphology at the CMB. To assess the 
agreement between the numerical models fields and the geomagnetic field at the 
CMB, we resort to using the four rating parameters introduced by 
\cite{Christensen10}.

To meet these main objectives, we conduct a systematic parameter study 
varying $\hstrat$ from 0 to 290~km and $N_m/\Omega$ from 0 to more than 50 for 
different combinations of Ekman, Rayleigh and magnetic Prandtl numbers. 
This work complements previous studies on the same topic that have assumed 
weaker stratfication degrees $N_m/\Omega < 5$ 
\citep{Olson17,Yan18,Christensen18}.

The paper is organised as follows. The details of the numerical geodynamo 
model and the control parameters are introduced in section~\ref{sec:models}. 
Section~\ref{sec:results} presents the numerical results, while
section~\ref{sec:geoph} describes the geophysical implications. We conclude 
with a summary of our findings in section~\ref{sec:concl}.

\section{Dynamo model}
\label{sec:models}

\subsection{Model equations and control parameters}

We consider a spherical shell of inner radius $r_i$ and outer radius $r_o$ 
filled with an incompressible conducting fluid of constant density $\rho$ 
which rotates at a constant frequency $\Omega$ about the $z$-axis. 
We adopt a dimensionless formulation of the magneto-hydrodynamic 
equations under the Boussinesq approximation. In the following, we employ the 
shell thickness $d=r_o-r_i$ as the 
reference length scale and the viscous diffusion time $d^2/\nu$ as the 
reference time scale. Velocity is expressed in units of $\nu/d$ and magnetic 
field in units of $\sqrt{\rho\mu\lambda \Omega}$, where $\mu$ is the magnetic 
permeability, $\nu$ is the kinematic viscosity and $\lambda$ is the magnetic 
diffusivity. The temperature scale 
is defined using the value of the gradient of the background temperature $T_c$ 
at the inner boundary $\left|\mathrm{d} T_c / \mathrm{d} r\right|_{r_i}$ 
multiplied by the lengthscale $d$.

The dimensionless equations that control the time evolution of the velocity 
$u$, the magnetic field $\vec{B}$ and the temperature perturbation $\vartheta$ 
are then expressed by
\begin{equation}
 \vec{\nabla}\cdot\vec{u}=0\,,\quad \vec{\nabla}\cdot\vec{B}=0\,,
 \label{eq:divu}
\end{equation}
\begin{equation}
\begin{aligned}
 \dfrac{\partial \vec{u}}{\partial t}+\vec{u}\cdot\vec{\nabla u} + \dfrac{2}{E} 
\vec{e_z}\times \vec{u} = & -\vec{\nabla} p + \dfrac{Ra}{Pr} g\,\vartheta\, 
\vec{e_r} \\ & +\dfrac{1}{E\,Pm}\left(\vec{\nabla}\times 
\vec{B}\right)\times{\vec{B}}+ \nabla^2\vec{u}\,,
\end{aligned}
\label{eq:NS}
\end{equation}
\begin{equation}
 \dfrac{\partial B}{\partial t} = 
\vec{\nabla}\times\left(\vec{u}\times\vec{B}\right) 
+ \dfrac{1}{Pm} \nabla^2 \vec{B}\,,
\label{eq:induction}
\end{equation}
\begin{equation}
 \dfrac{\partial \vartheta}{\partial t} 
+\vec{u}\cdot\vec{\nabla}\vartheta+ u_r \dfrac{\mathrm{d} T_c}{\mathrm{d} 
r} = \dfrac{1}{Pr} \nabla^2 \vartheta\,,
\label{eq:temp}
\end{equation}
where $p$ is the pressure, $\vec{e_r}$ is the unit vector in the 
radial direction and $g=r/r_o$ is the dimensionless gravity profile. 
The dimensionless set of equations (\ref{eq:divu}-\ref{eq:temp}) is governed 
by four dimensionless control parameters, namely the Ekman number $E$, the 
Rayleigh number $Ra$, the Prandtl number $Pr$ and the magnetic Prandtl number 
$Pm$ defined by
\begin{equation}
 E = \dfrac{\nu}{\Omega d^2},\, Ra = \dfrac{\alpha g_o d^4}{\nu \kappa} 
\left|\dfrac{\mathrm{d} T_c}{\mathrm{d} 
r}\right|_{r_i},\, Pr = \dfrac{\nu}{\kappa},\,Pm = \dfrac{\nu}{\lambda}\,,
\end{equation}
where $\alpha$ is the thermal expansivity, $g_o$ is the gravity at the outer 
boundary and $\kappa$ is the thermal diffusivity. 

The location and the degree of stratification of the stable layer 
are controlled by the radial variations of the gradient of the temperature 
background $\mathrm{d} T_c / \mathrm{d} r$. In regions where $\mathrm{d} T_c 
/\mathrm{d} r < 0$, the flow is indeed convectively-unstable, while 
stably-stratified regions correspond to $\mathrm{d} T_c /\mathrm{d} r > 0$.
We adopt here a simplified parametrised background temperature gradient 
to easily vary the location and the amplitude of 
the stably-stratified region. 

To do so, one possible approach, introduced by 
\cite{Takehiro01}, consists in assuming an homogeneous volumetric heat source 
in 
the convectively-unstable region and a constant positive temperature gradient 
$\mathrm{d} T_c /\mathrm{d} r$ in the stably-stratified outer layer. A 
continuous profile is then obtained by introducing a smooth $\tanh$ function 
centered at the transition radius $r_s$. This approach has the disadvantage of 
introducing  an additional parameter $\sigma$ which controls the stiffness of 
the transition between the two 
layers \citep[e.g.][]{Nakagawa11,Nakagawa15}.

A possible way out to remove the ambiguity of defining a suitable value for  
$\sigma$ consists in rather assuming that the degree of stratification grows 
linearly with radius accross the stably-stratified layer 
\citep[e.g.][]{Rieutord95,Lister98,Buffett14,Vidal15,Buffett16}. In this case, 
the maximum degree of stratification is reached at the CMB and linearly 
decreases to zero at the top of the convective part, in broad agreement with 
some seismic studies \citep[e.g.][]{Helffrich13}.  The temperature background 
$\mathrm{d}T_c/\mathrm{d}r$ is now entirely specified by the transition radius 
$r_s$ and the maximum degree of stratification $\Gamma$.
In the following, we adopt a  piecewise function defined by

\begin{equation}
 \dfrac{\mathrm{d} T_c}{\mathrm{d} r} = \left\lbrace
 \begin{aligned}
  -1, \quad r<r_s, \\
  \Gamma \dfrac{r-r_s}{\hstrat}+\dfrac{r-r_o}{\hstrat}, \quad r\geq r_s, \\
 \end{aligned}
 \right.
 \label{eq:dTdr}
\end{equation}
where $\hstrat=r_o-r_s$ corresponds to the thickness of the 
stable layer. The control parameter $\Gamma$  is related to the value of the 
Brunt-V\"ais\"al\"a frequency at the CMB $N_m$ via
\begin{equation}
 \dfrac{N_m}{\Omega} = \sqrt{\dfrac{Ra\,E^2}{Pr}\,\Gamma}\,.
 \label{eq:N_Omega}
\end{equation}

The set of equations (\ref{eq:divu}-\ref{eq:temp}) is supplemented by boundary 
conditions. We assume here rigid mechanical boundaries at both the 
ICB and the CMB. We employ mixed thermal boundary conditions with 
\[
 \left.\vartheta\right|_{r=r_i}=0,\quad \left.\dfrac{\partial 
\vartheta}{\partial r}\right|_{r=r_o}=0\,.
\]
This choice of thermal boundary conditions grossly reflects a 
fixed solidification temperature at the inner core boundary and a fixed flux 
extracted by the mantle at the CMB.
The magnetic field is matched to a potential field at the outer boundary, while 
the inner core is treated  as an electrically-conducting rigid 
sphere which is free to rotate about the $z$-axis.

\subsection{Numerical method}

The majority of the simulations computed in this study have been carried out 
using the open-source code \texttt{MagIC} \citep[][freely available 
at \url{https://github.com/magic-sph/magic}]{Wicht02}, while some complementary
simulations were integrated using the \texttt{PARODY-JA} code 
\citep{Dormy98,Aubert08}.

The set of equations (\ref{eq:divu}-\ref{eq:temp}) is solved in 
the spherical coordinates $(r,\theta,\phi)$ by expanding the velocity and the 
magnetic fields into poloidal and toroidal potentials

\[
 \begin{aligned}
 \vec{u} &=\vec{\nabla}\times\left(\vec{\nabla} \times W\,\vec{e_r}\right)+
\vec{\nabla}\times Z\,\vec{e_r}\,,\\
  \vec{B} &=\vec{\nabla}\times\left(\vec{\nabla} \times G\,\vec{e_r}\right)+
\vec{\nabla}\times H\,\vec{e_r}\,.
\end{aligned}
\]
The unknowns $W$, $Z$, $G$, $H$, $\vartheta$ and $p$ are expanded in spherical 
harmonic 
functions up to degree $\ell_{\text{max}}$ in the angular directions. In the 
radial direction, \texttt{MagIC} uses a Chebyshev collocation method with $N_r$ 
radial grid points $r_k$ defined by
\[
 r_k = \dfrac{1}{2} (x_k + r_o+r_i),\ x_k = 
\cos\left[\dfrac{(k-1)\pi}{N_r-1}\right],
\]
for $k\in[1,N_r]$, while \texttt{PARODY-JA} adopts a second-order finite 
difference scheme with $N_r$ grid points. For both codes, the equations are 
advanced in time using an implicit-explicit Crank-Nicolson second-order 
Adams-Bashforth scheme, which 
treats the nonlinear terms and the Coriolis force explicitly and the remaining 
terms implicitly.  The advection of the background temperature gradient $u_r 
\,\mathrm{d} T_c/\mathrm{d} r$ is handled implicitly when $N > \Omega$ to 
avoid severe time step limitations that would otherwise occur because of the 
propagation of gravity waves \citep[for a comparison, see][]{Brown12}.
\cite{Glatzmaier84}, \cite{Tilgner97} or \cite{Christensen15} 
provide a more comprehensive description of the numerical method and the 
spectral transforms involved in the computations. In both 
\texttt{MagIC} and \texttt{PARODY-JA}, the spherical transforms are handled 
using the open-source library \texttt{SHTns} \citep[][freely available at 
\url{https://bitbucket.org/nschaeff/shtns}]{Schaeffer13}

Standard Chebyshev collocation points such as the Gauss-Lobatto nodal 
points $x_k$ feature a typical grid spacing that decays with $N_r^{-2}$ close 
to 
the boundaries. In presence of a sizeable magnetic field, 
this imposes severe time step restrictions due to the propagation of Alfv\'en 
waves in the vicinity of the boundaries \citep[e.g.][]{Christensen99}. To 
alleviate this limitation, we adopt in \texttt{MagIC} the mapping from 
\cite{Kosloff93} defined by
\[
 y_k = \dfrac{\arcsin (\alpha_\text{map}\,x_k)}{\arcsin (\alpha_\text{map})},\ 
k=1, \cdots, N_r\,
\]
where $0 \leq \alpha_\text{map} < 1$ is the mapping coefficient.
This mapping allows a more even redistribution of the radial grid points 
\citep[see][ \S 16.9]{Boyd01}. To maintain the spectral convergence of the 
radial scheme, the mapping coefficient $\alpha_\text{map}$ has to be kept 
under a threshold value defined by
\[
 \alpha_{\text{map}} \leq \left[\cosh \left(\dfrac{|\ln\epsilon|}{N_r-1} \right)
\right]^{-1}
\]
where $\epsilon$ is the machine precision. Comparison of simulations with or 
without this mapping shows an increased average timestep size by a factor of 
two.

\subsection{Parameters choice and diagnostics}

A systematic parameter study has been conducted varying the Ekman 
number between $E=3\times 10^{-4}$ and $E=10^{-6}$, the Rayleigh 
number between $Ra=3\times 10^6$ and $Ra=9\times 10^{10}$ and the 
magnetic Prandtl number within the range $0.5<Pm<5$. For all the 
numerical models, $Pr$ is kept 
fixed to $1$. The influence of the stable layer has been studied by varying its 
degree of stratification within the range $0\leq N_m/\Omega<52$ and its 
thickness 
using the following values $\hstrat \in [0, 53, 87, 155, 200, 290]$~km.
Throughout the paper, the conversion between dimensionless and 
dimensional lengthscales is obtained by assuming $d=2260$~km.
To ensure a good statistical convergence, the numerical 
models have been integrated for at least half a magnetic diffusion time 
$\tau_\lambda$, except for the simulation with $E=10^{-6}$ which has 
been integrated over $0.2\,\tau_\lambda$.
In total, 70 direct numerical simulations detailed in 
Table~\ref{tab:results} 
have been computed in this study.

% \subsection{Diagnostics}

In the following, we employ overbars to denote time averages and angular 
brackets to express volume averages:
\[
 \langle f \rangle = \dfrac{1}{V}\int_V f\mathrm{d}V,\quad
 \overline{f}= \dfrac{1}{\tau}\int_{t_o}^{t_o+\tau} f \mathrm{d} t,
\]
where $V$ is the spherical shell volume, $t_o$ is the starting time for 
averaging and $\tau$ is the time-averaging period. The integration over a 
spherical surface is expressed by
\[
 \langle f \rangle_s = \int_0^\pi\int_{0}^{2\pi} f(r,\theta,\phi) \sin\theta 
\mathrm{d}\theta\mathrm{d}\phi\,.
\]

The typical flow amplitude is expressed by the magnetic Reynolds number $Rm$ 
defined by
\begin{equation}
 Rm = \overline{\langle u^2 \rangle^{1/2}}\,Pm\,,
\end{equation}
while the mean magnetic field amplitude is given by the Elsasser number 
$\Lambda$
\begin{equation}
 \Lambda = \overline{\langle B^2 \rangle}\,.
\end{equation}
To characterise the typical convective flow lengthscale, we introduce 
the mean spherical harmonic degree at the radius $r$
\[
 \bar{\ell}(r) = \overline{\dfrac{\sum_\ell \ell\,u_\ell^2(r)}{\sum_\ell
u^2_\ell(r)}},
\]
and the corresponding lengthscale
\[
 \mathcal{L}(r) = \dfrac{\pi\,r}{\bar{\ell}(r)},
\]
where $u^2_\ell(r)$ corresponds to the kinetic energy content at the 
spherical harmonic degree $\ell$ and at the radius $r$
\citep{Christensen06a}. In the following we will mainly focus on the 
convective flow lengthscale at the transition between the stably-stratified 
outer layer and the inner convective core, denoted by
\begin{equation}
\mathcal{L}_s= \dfrac{\pi\,r_s}{\bar{\ell}_s}, \quad
 \bar{\ell}_s = \bar{\ell}(r=r_s)\,.
\end{equation}

The morphological agreement between the magnetic fields produced in the 
numerical models and the geomagnetic field is assessed by four criteria 
introduced by \cite{Christensen10}. This involves physical quantities defined 
using the spectral properties of the magnetic field at the CMB for 
spherical harmonic degree and order lower than 8. The 
ratio of power between the axial dipole and the non-dipolar contributions 
defines the parameter AD/NAD. The degree of equatorial symmetry of the CMB 
field is measured by the parameter O/E, while the ratio of power between the 
axisymmetric and the non-axisymmetric contributions for the 
non-dipolar field is given by Z/NZ. Finally, the magnetic flux 
concentration factor FCF is defined by the variance of the square of the radial 
component of the magnetic field at the CMB. The combination of the time-average 
of these four quantities allow to estimate the degree of compliance  $\chi^2$ 
between the numerical model field and the geomagnetic field \citep[see][for the 
details]{Christensen10}. 

Table~\ref{tab:results} summarises the values of the main diagnostics for all 
the simulations computed in this study.

% The buoyancy power is expressed by
% 
% \begin{equation}
%  
% \end{equation}
% where 
% \[
%  \langle f \rangle_S = \int_S f \mathrm
% \]

\section{Results}
\label{sec:results}

\subsection{Penetrative rotating convection}

\begin{figure*}
 \centering
 \includegraphics[width=16cm]{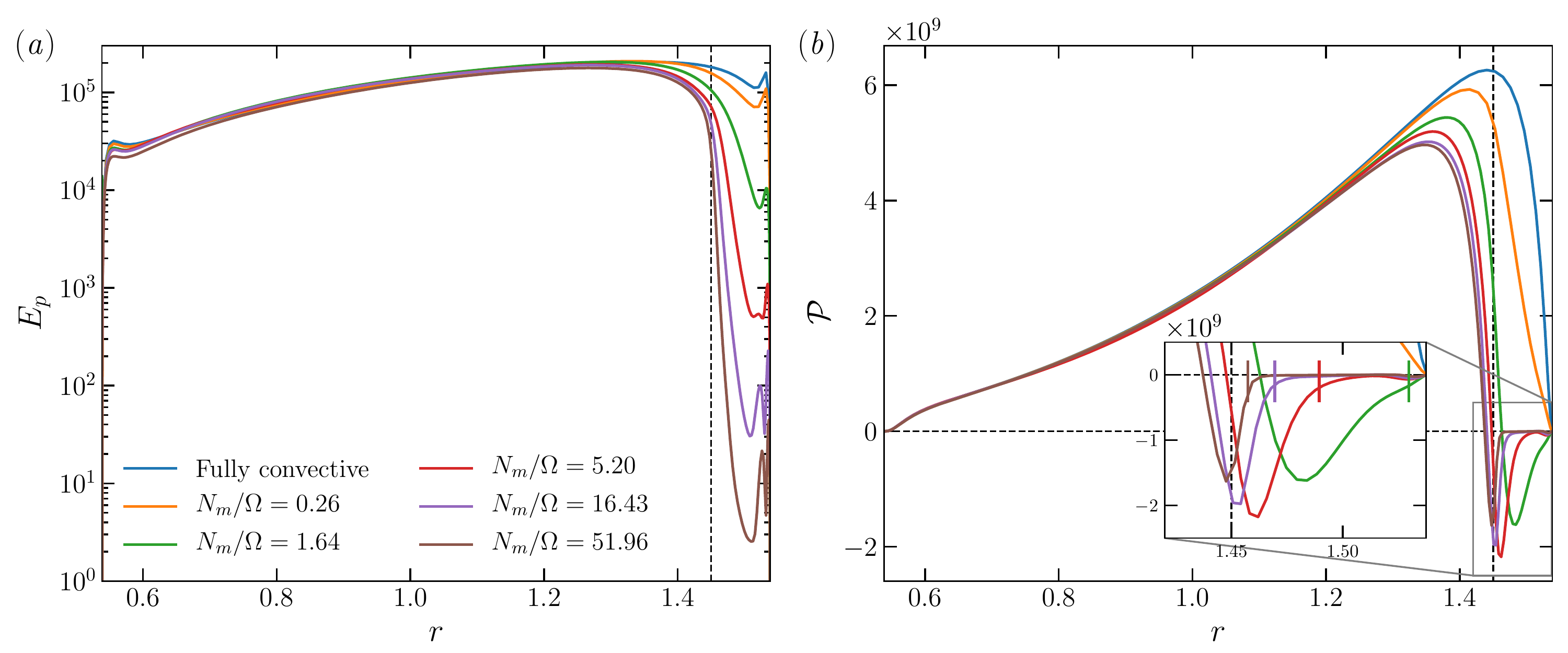}
 \caption{(\textit{a}) Time-averaged poloidal kinetic energy 
(Eq.~\ref{eq:ekpol}) as a function of 
radius for numerical models with $E=3\times 10^{-5}$, $Ra=3\times 
10^{8}$, $Pm=2.5$, $r_s=1.45$ ($\hstrat=200$~km) and different values of 
$N_m/\Omega$. The vertical dashed line corresponds to $r=r_s$. (\textit{b}) 
Time-averaged buoyancy power $\mathcal{P}$ (Eq.~\ref{eq:power}) as a function 
of radius. The vertical dashed line corresponds to $r=r_s$, while 
the horizontal dashed line corresponds to the neutral buoyancy line 
$\mathcal{P}=0$. The zoomed-in inset highlights the radial profiles of 
$\mathcal{P}$ in the stably-stratified layer. The small colored vertical 
segments mark the extent of the convective penetration $r_p$ defined in 
Eq.~(\ref{eq:rp}).}
 \label{fig:ekin_buo_r}
\end{figure*}

A stably-stratified layer lying above a convective region does not act as a 
simple rigid wall that would quench all convective motions. In practice, the 
parcels of fluid which are moving outward in the vicinity of the interface 
rather penetrate over some distance $\dpen$ into the stably-stratified layer, 
gradually loosing their momentum. An easy and practical way to visualise 
this phenomenon \citep[e.g.][]{Rogers05} resorts to looking at the radial 
profile of poloidal kinetic energy averaged over time
\begin{equation}
 E_p = \frac{1}{2}\sum_{\ell, m} 
\ell(\ell+1)\left[
   \frac{\ell(\ell+1)}{r^2}\overline{|W_{\ell 
m}|^2}+\overline{\left|\frac{{\rm 
d} W_{\ell m}}{{\rm 
d} r}\right|^2}   \right]\,,
\label{eq:ekpol}
\end{equation}
where $W_{\ell m}$ is the poloidal potential at degree $\ell$ and order $m$.
Figure~\ref{fig:ekin_buo_r}\textit{a} shows the comparison of $E_p$ for one 
fully-convective model and for five simulations with $\hstrat=200$~km and an
increasing degree of stratification $N_m/\Omega$. All models exhibit comparable 
profiles in most of the convective core and only start to depart from each 
other in the upper part of the convective region. In the stably-stratified 
outer layer, the poloidal energy content decreases with increasing values of 
$N_m/\Omega$. While the simulation with $N_m/\Omega=0.26$ is comparable to the 
fully-convective model in this region, the case with the strongest 
stratification $N_m/\Omega=51.96$ features an energy content roughly four orders 
of magnitude below its fully convective counterpart.

The radial profiles of $E_p$ can be further employed to estimate the distance 
of penetration $\dpen$ either by measuring the point where $E_p$ 
drops below a given fraction of its maximum value \citep[e.g.][]{Rogers05}, or 
by measuring the $e$-folding distance of $E_p$ at the edge of the convective 
layer \citep[e.g.][]{Takehiro01}. Both methods carry their 
own limitations: the former is very sensitive to the threshold value when $E_p$ 
shows a stiff decay at the transition; while the latter can yield $\dpen$ 
larger than the actual thickness of the stably-stratified layer 
\citep[see][Fig.~10]{Dietrich18}.

A complementary approach, which has proven to be insightful in the context 
of Solar convection  \citep[e.g.][]{Browning04,Deng08,Brun17}, resorts to 
studying the radial variations of the convective flux or of the buoyancy power 
\citep[see][]{Takehiro18a} expressed by

\begin{equation}
 \mathcal{P} = \dfrac{Ra\,E}{Pr} g\,\overline{\langle u_r \vartheta 
\rangle_s}\,.
 \label{eq:power}
\end{equation}
Figure~\ref{fig:ekin_buo_r}\textit{b} shows the radial profiles of 
$\mathcal{P}$ for the same numerical simulations as in 
Fig.~\ref{fig:ekin_buo_r}\textit{a}. In the convective core, the 
eddies which are hotter (colder) than their surroundings are moving outward 
(inward), yielding a positive buoyancy power $\mathcal{P}$.
But when a convective 
parcel overshoots in the sub-adiabatic layer, the positive radial 
velocity becomes anti-correlated with the negative
thermal fluctuations, yielding $\mathcal{P} < 0$ at the base of the 
stably-stratified layer \citep[e.g.][]{Takehiro18a}.
As shown in the 
inset of Fig.~\ref{fig:ekin_buo_r}\textit{b}, the radial extent of the fluid 
region where $\mathcal{P}<0$ is a decreasing function of $N_m/\Omega$.  
Following \cite{Browning04}, the upper boundary of the overshooting region can 
be defined by the radius at which the buoyancy power attains 10\% of its 
minimum negative value
\begin{equation}
\mathcal{P}(r_p) = 0.1\min(\mathcal{P}) \quad\text{and}\quad r_p > r_\text{min},
\label{eq:rp}
\end{equation}
where $r_\text{min}$ corresponds to the radius where the buoyancy power 
reaches its minimum. This definition still involves an arbitrary threshold 
value, but  $r_p$ has been found by previous studies to be fairly insensitive 
to this \citep[e.g.][]{Brun11}. The location of $r_p$ using this definition are 
marked by vertical segments in Fig.~\ref{fig:ekin_buo_r}\textit{b}. We then 
define the penetration depth $\dpen$ by
\begin{equation}
 \dpen=r_p-r_s\,.
 \label{eq:dpen}
\end{equation}
The adopted definition of $r_p$ guarantees that the penetration depth 
remains bounded by $\hstrat$, i.e. $\max(\dpen) \leq \hstrat$.

\begin{figure*}
 \centering
 \includegraphics[width=16cm]{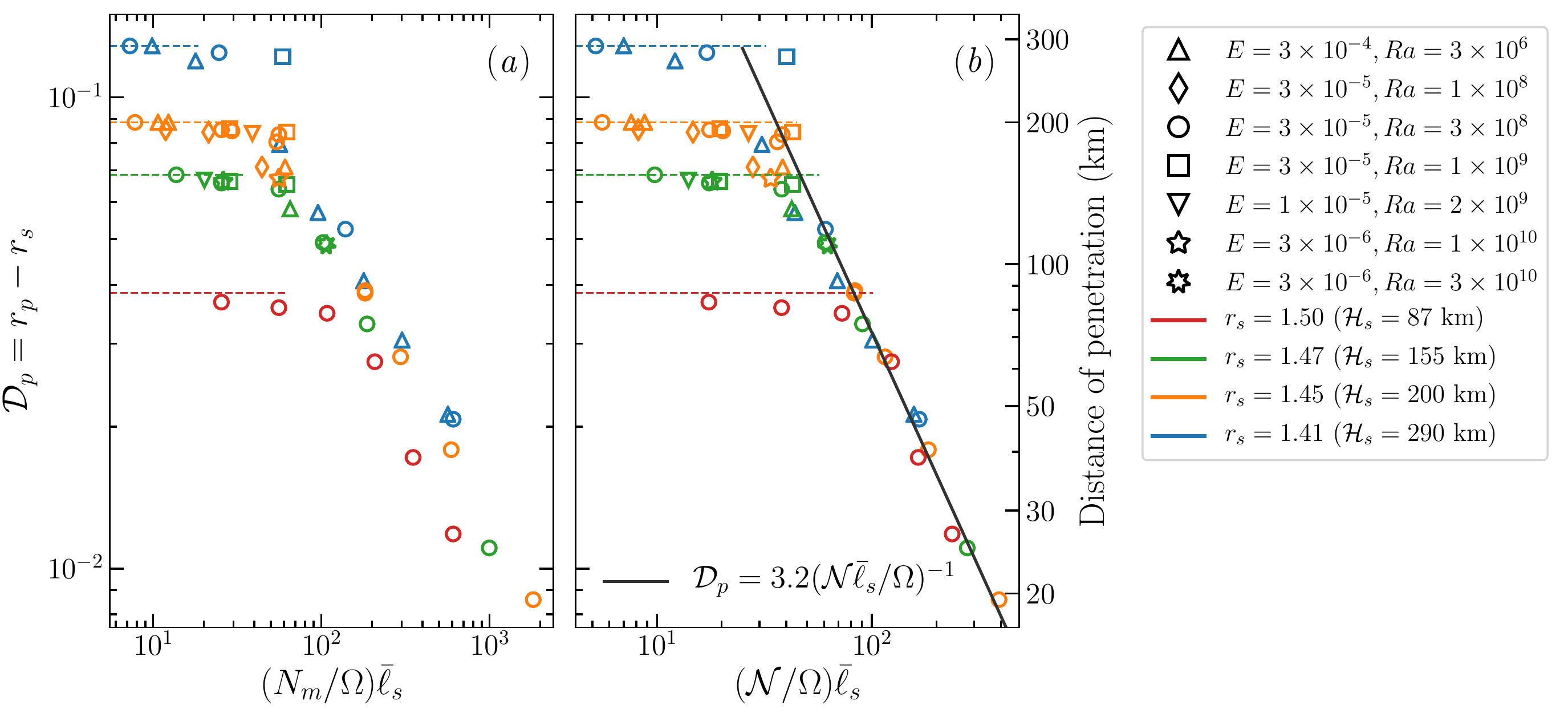}
 \caption{Distance of penetration of the convective flow $\dpen$ 
(Eq.~\ref{eq:dpen}) as a function 
of $(N_m/\Omega)\,\bar{\ell}_s$ (left panel) and as a function of 
$(\mathcal{N}/\Omega)\,\bar{\ell}_s$ (right panel). The color of the symbols 
correspond to the thickness of the stratified layer $\hstrat$, while the shape 
correspond to different $(E,Ra)$ combination of parameters listed in 
Tab.~\ref{tab:results}. In each panel, the colored dashed lines correspond 
to the maximum extent of the penetration, i.e. $\dpen=\hstrat$. The solid black 
in panel (\textit{b}) line corresponds to a best fit for the models with 
$(\mathcal{N}/\Omega)\,\bar{\ell}_s > 80$.}
 \label{fig:penet}
\end{figure*}

We now examine how $\dpen$ evolves with the degree of stratification 
$N_m/\Omega$. In the physical regime of rapidly-rotating convection and in 
absence of magnetic field, the linear 
stability analysis by \cite{Takehiro01} suggest that the distance of 
penetration is inversely proportional to the ratio of the 
Brunt-V\"ais\"al\"a and the rotation frequencies (Eq.~\ref{eq:dpen_takehiro}).
It is however not entirely clear whether this scaling should still hold 
in presence of a magnetic field \citep{Takehiro15}.

Figure~\ref{fig:penet}\textit{a} shows $\dpen$ as a function of  
$(N_m/\Omega)\bar{\ell}_s$ for all the numerical simulations computed in this 
study. For each stable layer thickness $\hstrat$, the evolution of $\dpen$ with 
$(N_m/\Omega)\bar{\ell}_s$ is comprised of two parts: one nearly horizontal 
part where the degree of stratification is weak enough such that $\dpen\simeq 
\hstrat$; and a second branch for $(N_m/\Omega)\bar{\ell}_s > 100$ where 
$\dpen$ decreases with the degree of stratification. However, a 
dependence to $\hstrat$ is still visible in the decaying branch. At a fixed 
value of $(N_m/\Omega)\bar{\ell}_s$, the penetration distance can indeed vary 
by a factor of roughly two \citep[see also][]{Dietrich18}. We attribute this 
remaining dependence to the local radial variations of the Brunt-V\"ais\"al\"a 
frequency (Eq.~\ref{eq:dTdr}). Since the degree of stratification almost 
linearly increases from neutral stability at the edge of the convective layer 
to $N_m/\Omega$ at the CMB, a convective eddy that penetrates deep in the 
stable layer does not feel the same stratification as one that would hardly 
scratch into it. To account for this effect, we introduce an \emph{effective 
stratification} $\mathcal{N}/\Omega$ defined by the averaged 
Brunt-V\"ais\"al\"a frequency between the spherical shell radii $r_s$ and $r_p$

\begin{equation}
 \left(\dfrac{\mathcal{N}}{\Omega}\right)^2 = \dfrac{RaE^2}{Pr}
\dfrac{\displaystyle\int_{r_s}^{r_p} g\dfrac{\mathrm{d} 
T_c}{\mathrm{d} r} r^2\mathrm{d}r}{%
  \displaystyle\int_{r_s}^{r_p} r^2 \mathrm{d}r}\,.
\label{eq:effN_Om}
\end{equation}
Figure~\ref{fig:penet}\textit{b} shows $\dpen$ as a function of 
$(\mathcal{N}/\Omega)\bar{\ell}_s$. In contrast to
Fig.~\ref{fig:penet}\textit{a}, the measured penetration distances $\dpen$ now 
collapse on one single scaling behaviour. A best fit for the 
strongly-stratified simulations with $(\mathcal{N}/\Omega)\,\bar{\ell}_s > 80$ 
yield
\begin{equation}
 \dpen = (3.19\pm0.67 
) \left(\dfrac{\mathcal{N}}{\Omega}\bar{\ell}_s\right)^{-1.00\pm0.04}
 \label{eq:fitpenet}
\end{equation}
in excellent agreement with the theoretical 
scaling (\ref{eq:dpen_takehiro}) from \cite{Takehiro01}.

\begin{figure*}
 \centering
 \includegraphics[width=15cm]{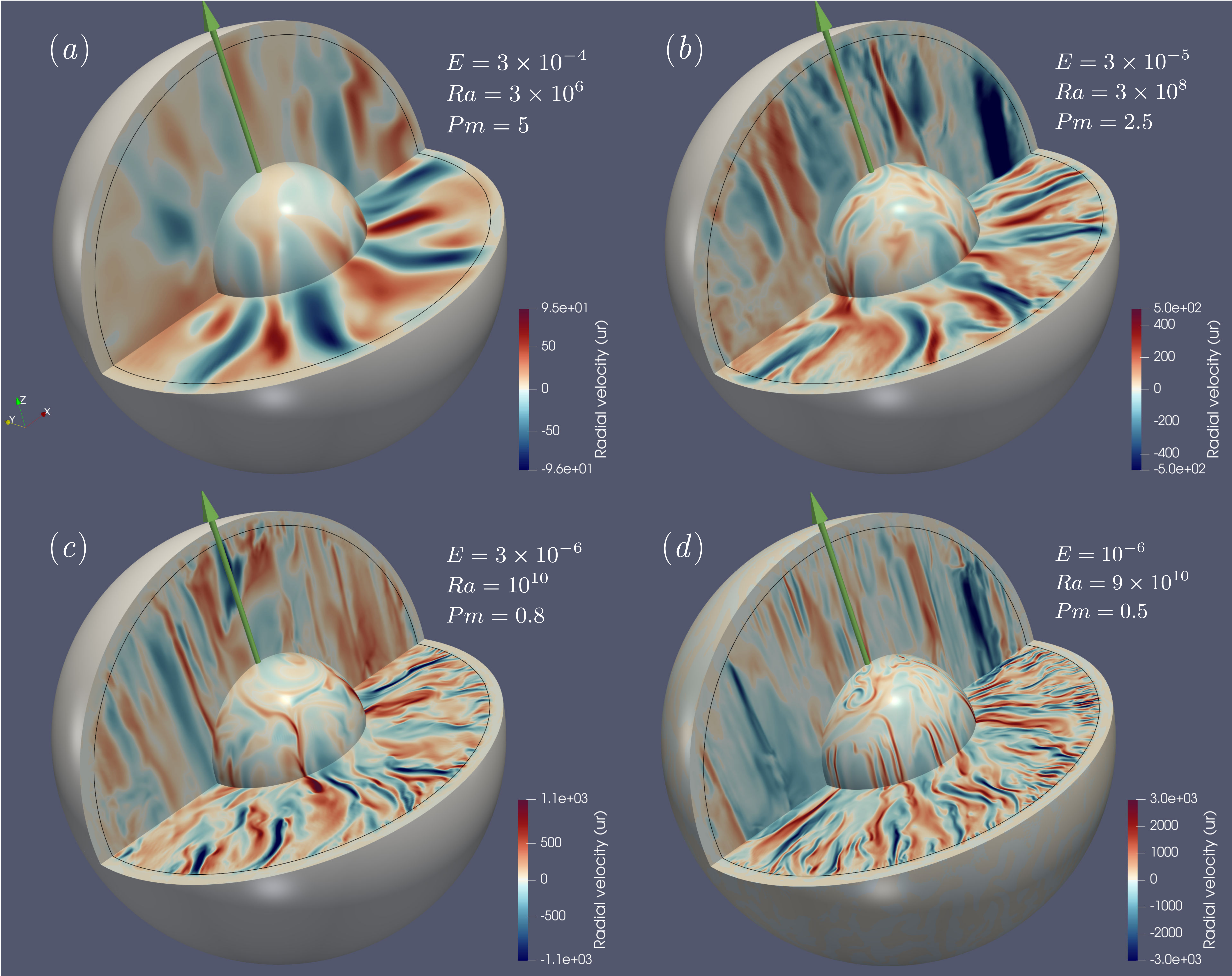}
 \caption{3-D renderings of the radial velocity $u_r$ for four dynamo 
models with the same stably-stratified layer thickness ($r_s=1.45$, 
$\hstrat=200$~km) and degree of stratification $N_m/\Omega \simeq 0.94$.
For each panel, the solid lines delineates the radius of the stratified 
layer $r=r_s$, the green arrow highlights the rotation axis and the inner 
spherical surface corresponds to $r=0.39\,r_o$.}
 \label{fig:snaps}
\end{figure*}

Although this scaling has been theoretically derived in absence of magnetic 
field, the penetration distance of convective eddies in dynamo models is found 
to still only depend on the ratio of the local Brunt-V\"ais\"al\"a frequency to 
the rotation rate and on the typical horizontal size of the convective flow at 
the transition radius. This implies that at a given stratification degree, 
small scale eddies will penetrate over a shorter distance than the large ones.
To illustrate this physical phenomenon, Fig.~\ref{fig:snaps} shows snapshots of 
the radial component of the convective flow $u_r$ for four numerical 
simulations with comparable $N_m/\Omega$ but decreasing Ekman numbers from 
$E=3\times 10^{-4}$ (\textit{a}) to $E=10^{-6}$ (\textit{d}). 
The typical convective flow lengthscale in the upper part of the convective 
region decreases with the Ekman number and the penetration distance decreases 
accordingly. For the two cases with the lowest Ekman number, we observe a clear
separation between larger flow lengthscales in the bulk of the convective core 
and smaller scale features at $r_s$.
To quantify this scale separation, we thus introduce another 
lengthscale measure deeper in the convective region, denoted by
\begin{equation}
\mathcal{L}_b= \dfrac{\pi\,r_b}{\bar{\ell}_b}, \quad
 \bar{\ell}_b = \bar{\ell}(r=r_b)\,,
\end{equation}
where $r_b=r_i+0.25$.

\begin{figure*}
  \centering
 \includegraphics[width=16cm]{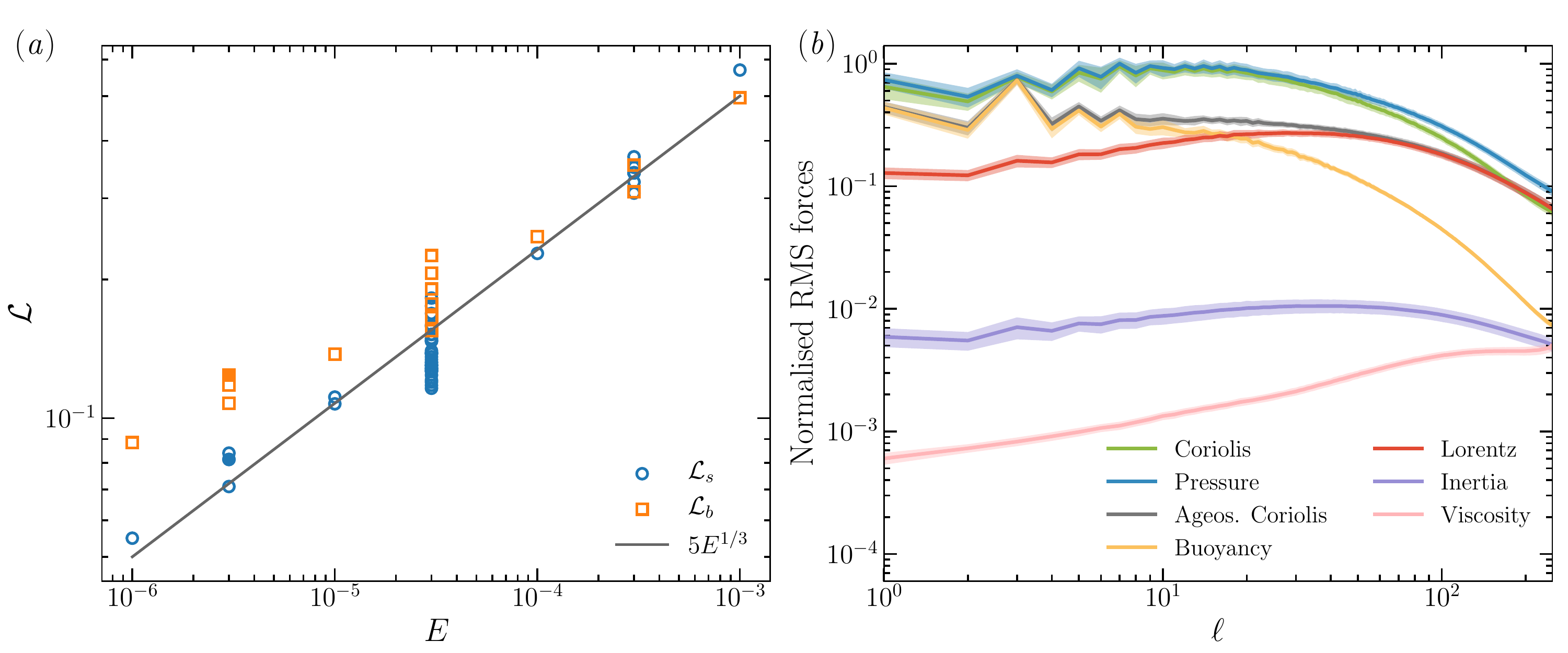}
 \caption{(\textit{a}) Time-averaged convective flow lengthscale at 
$r_s$ (i.e. $\mathcal{L}_s$)  and at $r_b$ (i.e. $\mathcal{L}_b$) as a function 
of the Ekman number $E$ for all the numerical models with a stratified layer 
(i.e. $\Gamma > 0$). (\textit{b}) Force balance spectra in the fluid 
bulk (Eq.~\ref{eq:forcebal}) normalised with 
respect to the maximum of the pressure force for a 
numerical simulation with $E=3\times 10^{-6}$, $Ra=10^{10}$, $Pm=0.8$, 
$r_s=1.45$ (i.e. $\hstrat=200$~km) and $N_m/\Omega=0.95$. The filled symbols in 
panel (\textit{a}) correspond to the model described in panel (\textit{b}).}
 \label{fig:forcebal}
\end{figure*}

Figure~\ref{fig:forcebal}\textit{a} shows $\mathcal{L}_s$ and 
$\mathcal{L}_b$ as a function of the Ekman number for all the numerical models 
that feature a stably-stratified layer (i.e. $\Gamma >0$). At the transition 
radius $r_s$, the convective flow lengthscale is found to follow a 
$\mathcal{L}_s \sim E^{1/3}$ law (solid line). This scaling reflects the 
local onset of convection beneath $r_s$ where the available power content 
drops and yields weaker local convective supercriticality (see 
Fig.~\ref{fig:ekin_buo_r}). The situation differs in the 
bulk of the convective core: while the flow lengthscale at $\mathcal{L}_b$ 
is almost identical to $\mathcal{L}_s$ when $E \geq 3\times 10^{-5}$, the two 
lengthscales gradually depart from each other at lower Ekman numbers with 
$\mathcal{L}_b > \mathcal{L}_s$. This confirms the 
scale separation observed in the numerical simulations with the lowest Ekman 
numbers shown in the lower panels of Fig.~\ref{fig:snaps}.

The deviation from the viscous scaling 
$\mathcal{L}_b \sim E^{1/3}$ \citep[e.g.][]{KingBuffett13,Gastine16} 
indicates that the underlying force balance which controls the convective flow 
is not dominated by viscous effects. Following \cite{Aubert17} and 
\cite{Schwaiger19}, we analyse this force balance by decomposing each term
that enter the Navier-Stokes equation (\ref{eq:NS}) into spherical harmonics
\begin{equation}
 F_{rms}^2 = \dfrac{1}{V}\int_{r_i+\lambda}^{r_o-\lambda}\sum_{\ell,m} 
F^2_{\ell m} r^2 \mathrm{d}r = \sum_\ell F_\ell^2\,,
\label{eq:forcebal}
\end{equation}
where $\lambda$ is the viscous boundary layer thickness. 
Figure~\ref{fig:forcebal}\textit{b} illustrates the normalised force balance 
spectra in the fluid bulk for a selected numerical simulation with 
$E=3\times 10^{-6}$, 
$Ra=10^{10}$, $r_s=1.45$ (i.e. $\hstrat=200$~km) and $N_m/\Omega=0.95$  using 
$\lambda =10^{-2} d$.
The leading order consists of a quasi-geostrophic (QG) force balance between 
Coriolis and pressure gradient. The  ageostrophic 
Coriolis contribution which accounts for the difference between Coriolis and 
pressure forces, is then equilibrated by buoyancy at large scales 
and by Lorentz force at small scales.  Inertia and viscosity lay one to two 
orders of magnitude below this second-order force balance.
This force hierarchy forms the so-called \emph{QG-MAC balance}
introduced by \cite{Davidson13}. This second-order force balance has 
been theoretically analyzed in the plane layer geometry by \cite{Calkins18} 
using a multiscale expansion and was reported in direct numerical 
simulations in spherical geometry
\citep[e.g.][]{Yadav16,Schaeffer17,Aubert17}. The force balance obtained 
in the numerical models with a $200$~km-thick stably-stratified layer atop 
the core is thus structurally akin to the force balance spectra of the fully 
convective simulations \citep[e.g.][]{Schwaiger19}.

\subsection{Skin-effect and magnetic field smoothing}

\begin{figure*}
 \centering
\includegraphics[width=16cm]{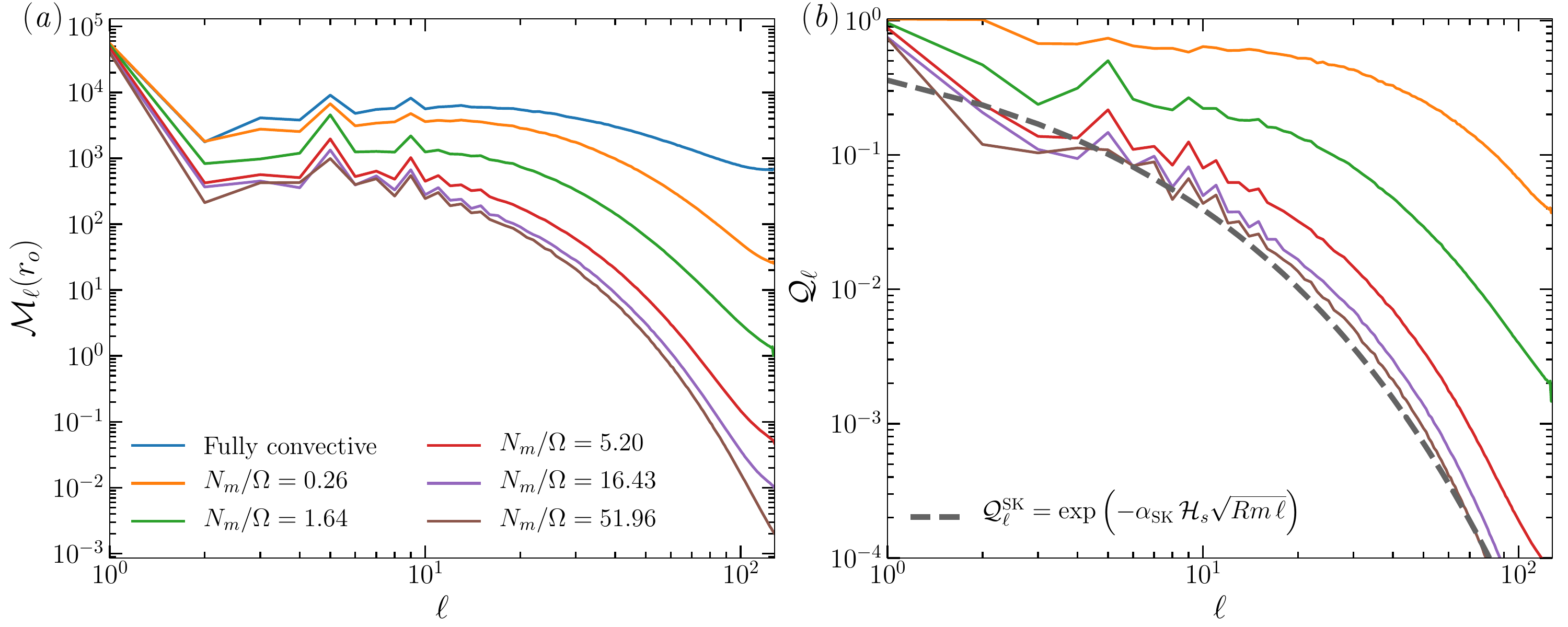}
 \caption{(\textit{a}) Time-averaged magnetic energy at the CMB 
$\ecmb$ as a function of the spherical harmonic degree $\ell$ 
for numerical models with $E=3\times 10^{-5}$, $Ra=3\times 10^{8}$, $Pm=2.5$, 
$r_s=1.45$ (i.e.  $\hstrat=200$~km) and increasing values of 
$N_m/\Omega$ (same models as in Fig.~\ref{fig:ekin_buo_r}). (\textit{b}) 
Damping of the magnetic energy at the CMB relative 
to the fully convective case $\mathcal{Q}_\ell$ (Eq.~\ref{eq:damping}) as a 
function of $\ell$. The dashed grey line corresponds to the scaling
Eq.~(\ref{eq:dampingmax}) using $\ask=0.5$ and the time-averaged magnetic 
Reynolds number of the fully convective case, i.e. $Rm=536$.}
 \label{fig:compSpec}
\end{figure*}

We now turn to examining the effect of the stable layer on the magnetic field 
structure. If one crudely assumes that the stable 
region is devoid of any fluid motion, it can be approximated by a layer of 
thickness $\hstrat$ filled with an electrically-conducting stagnant fluid. This 
then acts as a \emph{skin layer} that will attenuate the magnetic field 
amplitude by a factor $\exp(-\hstrat/\delta)$, where $\delta$ is the magnetic 
skin depth defined by
\[
 \delta \sim \sqrt{\dfrac{\tau_\ell}{Pm}}\,,
\]
where curvature effects due to spherical geometry have been neglected.
In the above expression, $\tau_\ell$ corresponds to the typical turnover time
$\tau_\ell \sim \mathcal{L}_s/Re$, where $Re=Rm/Pm$ is the fluid Reynolds 
number. This 
yields
\begin{equation}
\delta \sim  (Rm\,\bar{\ell}_s)^{-1/2}\,.
\label{eq:deltaskin} 
\end{equation}
The factor of attenuation of the magnetic energy due to the skin effect can 
hence be approximated by
\begin{equation}
 \ln\left[\dfrac{\mathcal{M}_\ell(r_o)}{\mathcal{M}_\ell(r_s)}\right] \sim 
-\hstrat (Rm\,\bar{\ell}_s)^{1/2}\,,
\label{eq:emagdamping}
\end{equation}
where $\mathcal{M}_\ell(r)$ corresponds to the magnetic energy at the spherical 
harmonic degree $\ell$ and at the radius $r$. From a practical stand-point, it 
is more convenient to assess the impact of a stable layer by a direct 
comparison of the magnetic energy at the CMB between a stably-stratified 
case and its fully convective counterpart
\begin{equation}
 \mathcal{Q}_\ell = 
\dfrac{\mathcal{M}_\ell^\text{strat}(r_o)}{\mathcal{M}_\ell^\text{FC}(r_o)}\,,
 \label{eq:damping}
\end{equation}
where the superscripts ``FC'' and ``strat'' stand for 
the fully convective and the stably-stratified models, respectively.
To relate the above expression to the skin effect (\ref{eq:emagdamping}),
we make the two following hypotheses:
  \begin{itemize}
      \item[$\bullet$] We assume that the magnetic energy at the transition 
radius $r_s$ is independent of the presence of a stable layer, i.e. 
$\mathcal{M}_\ell^\text{strat}(r_s) \simeq 
\mathcal{M}_\ell^\text{FC}(r_s)$,
   \item[$\bullet$] We assume that the magnetic energy of the fully convective 
model at $r_s$ is comparable to the energy at the CMB, i.e. 
$\mathcal{M}_\ell^{FC}(r_s) \simeq \mathcal{M}_\ell^{FC}(r_o)$.
  \end{itemize}
The validity of those hypotheses will be further assessed below.
Combining Eq.~(\ref{eq:emagdamping}) with the two previous assumptions yields 
the following scaling for the damping factor
\begin{equation}
 \qsk = \exp\left[-\ask \hstrat (Rm\,\bar{\ell}_s)^{1/2} \right]\,,
\label{eq:dampingmax}
\end{equation}
where $\ask$ is a proportionality coefficient that depends on 
the geometry. The above scaling should be understood as the 
maximum damping that a stable layer could yield in the idealised limit of 
vanishing fluid motions there, i.e. 
$\sup({\mathcal{Q}_\ell})=\qsk$ when $N_m/\Omega \gg 1$.

\begin{figure*}
 \centering
  \includegraphics[width=15cm]{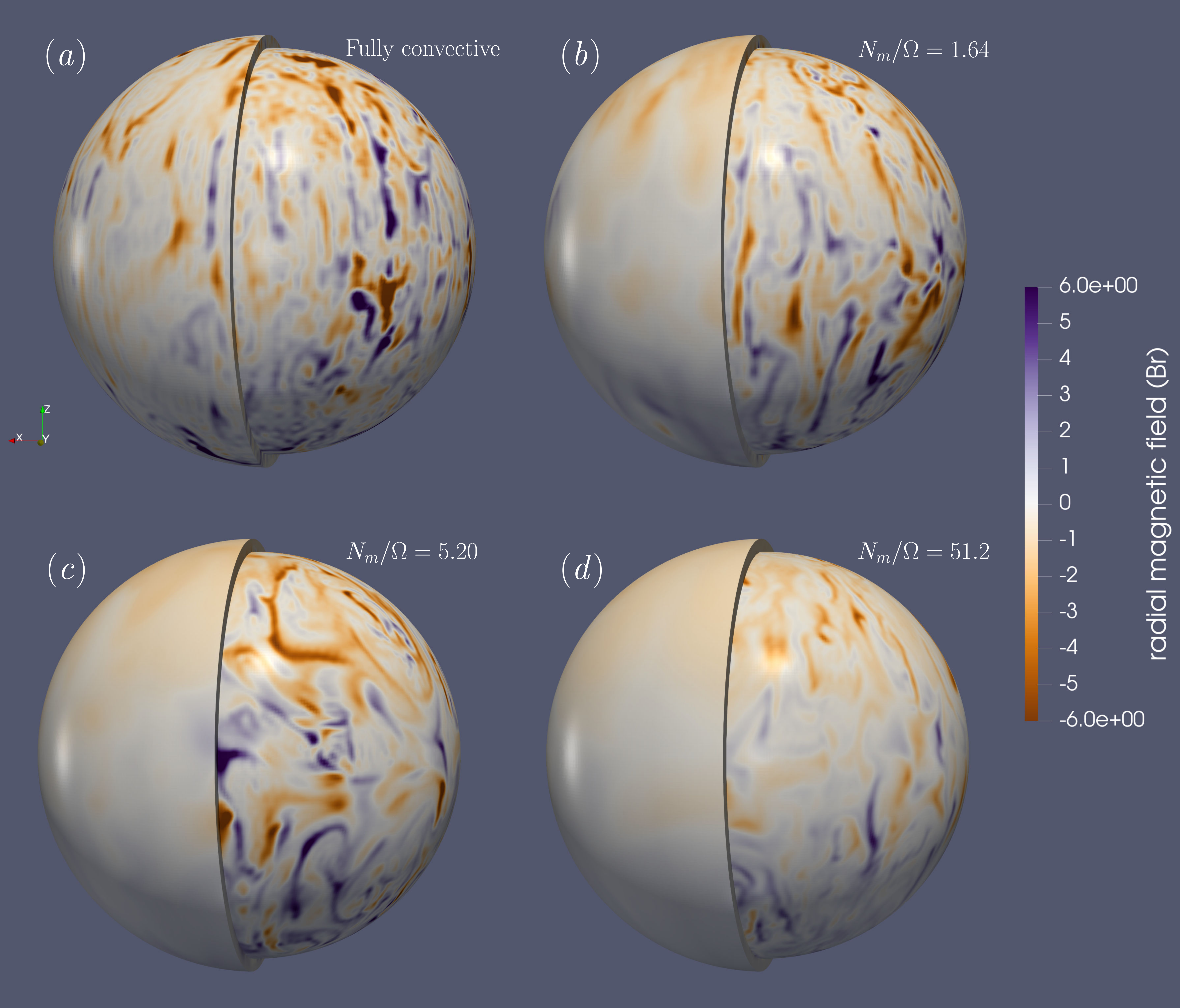}
  \caption{3-D renderings of the radial component of the magnetic field $B_r$ 
for four numerical models with the same control parameters $E=3\times 10^{-5}$, 
$Ra=3\times 10^8$, $Pm=2.5$ and increasing degree of stratification 
$N_m/\Omega$. The stratified cases have $r_s=1.45$ (i.e. 
$\hstrat=200$~km). The inner spheres correspond to $r=r_s$  and the outer ones 
to the CMB. The magnetic field amplitude is expressed in units of the square 
root of the Elsasser number.}
\label{fig:BrSmoothing}
\end{figure*}

Figure~\ref{fig:compSpec} shows the time-averaged magnetic energy spectra at 
the CMB (panel \textit{a}) and the damping factor $\mathcal{Q}_\ell$ (panel 
\textit{b}) for one fully convective simulation and five numerical models with 
an increasing degree of stratification $N_m/\Omega$ (same models as in 
Fig.~\ref{fig:ekin_buo_r}). The magnetic energy content decreases when 
increasing $N_m/\Omega$. This energy drop is more pronounced for the 
smaller scales of the magnetic field. A saturation is observed for 
the models with $N_m/\Omega > 10$ for which the spectra become 
comparable. The damping factor $\mathcal{Q}_\ell$ drops accordingly when 
increasing $N_m/\Omega$ to tend towards the limit $\qsk$, obtained here using 
the value of $Rm$ of the fully convective simulation and $\ask=0.5$ (dashed 
line in Fig.~\ref{fig:compSpec}\textit{b}). This implies that for large degree 
of stratification $N_m/\Omega \gg 1$, a stable layer has a similar dynamical 
signature on the magnetic field as a passive conductor of the same 
thickness. This is not the case for intermediate stratification $N_m/\Omega 
\simeq 1$ for which convective motions can penetrate into the stable layer over 
some distance $\dpen$.

To further illustrate the magnetic field damping due to the presence of a 
stable layer, Figure~\ref{fig:BrSmoothing} shows snapshots of the radial 
component of the magnetic field at the radius $r_s$ and at the CMB, for one 
fully-convective model and three simulations with increasing $N_m/\Omega$.
At the transition radius $r_s$, the magnetic field structures of the four 
cases are relatively similar, featuring a dominant dipolar structure 
accompanied by intense localised flux concentration. The first hypothesis 
involved in the derivation of Eq.~(\ref{eq:dampingmax}) is hence roughly 
satisfied, though a small decay of magnetic field amplitude with $N_m/\Omega$ 
is visible. This can be likely attributed to the decreasing available 
buoyancy power in the upper regions of the convective part (see 
Fig.~\ref{fig:ekin_buo_r}\textit{a}). For the fully-convective simulation, the 
magnetic field structure remains very similar at the CMB, validating the second 
assumption used when deriving Eq.~(\ref{eq:dampingmax}). In contrast, the 
stably-stratified layer reduces the magnetic field amplitude and acts as a 
low-pass filter on the magnetic field structures gradually filtering out the 
small-scale features when $N_m/\Omega$ increases. While inverse polarity 
patches are for instance still discernible on the $N_m/\Omega = 1.64$ case 
(Fig.~\ref{fig:BrSmoothing}\textit{b}), they disappear completely in the most 
stratified case with $N_m/\Omega = 51.2$ (Fig.~\ref{fig:BrSmoothing}\textit{d}).
We can hence anticipate that large degree of stratification will yield 
smooth CMB magnetic fields incompatible with the observed geomagnetic field 
\citep[see][]{Christensen18}.

\subsection{Earth-likeness}

\begin{figure*}
 \centering
 \includegraphics[width=17cm]{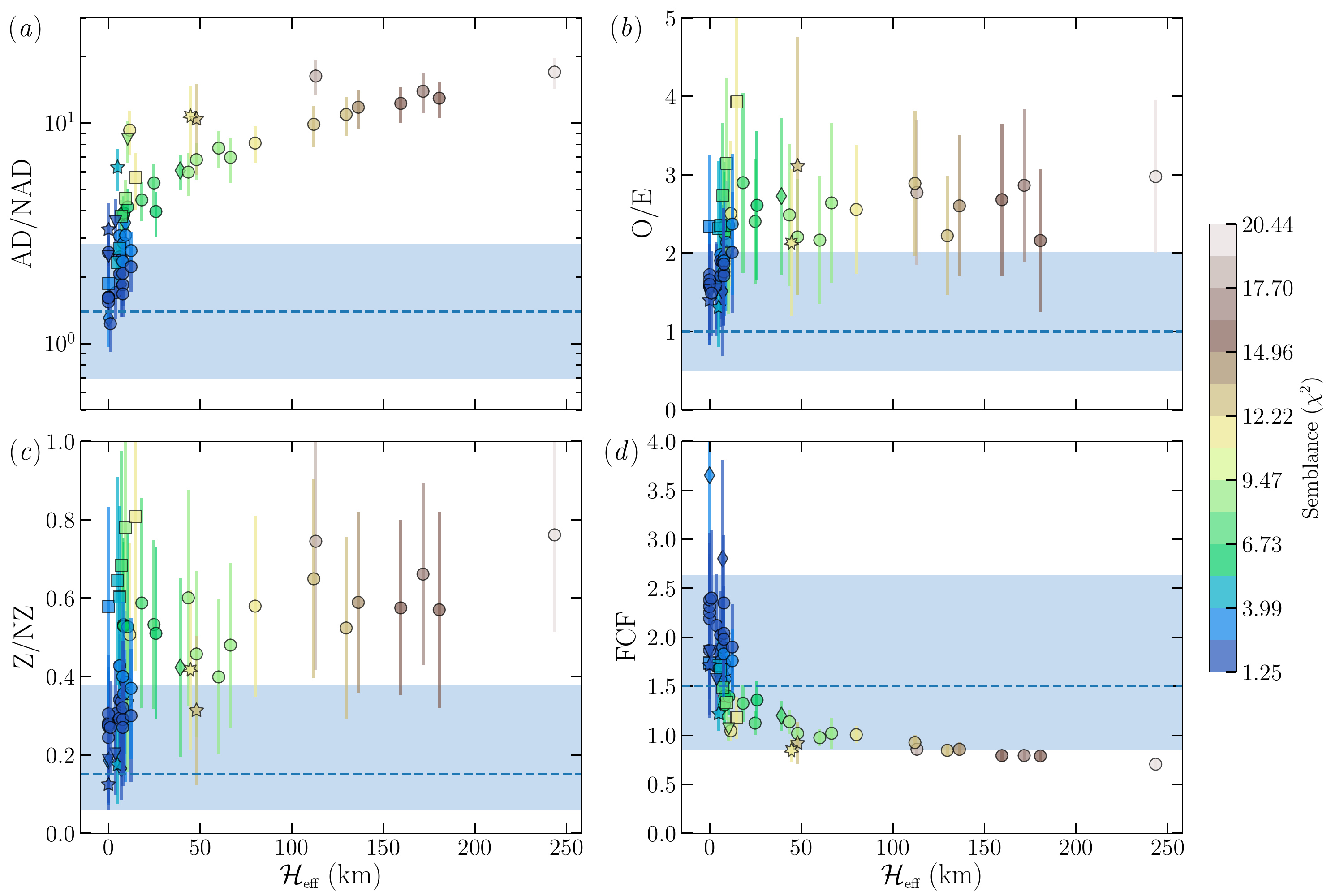}
 \caption{(\textit{a}) AD/NAD as a function of  the effective thickness of the 
stably stratified layer $\heff=r_o-r_p$. (\textit{b}) O/E as a function of 
$\heff$. (\textit{c}) Z/NZ as a function of $\heff$. (\textit{d}) FCF as a 
function of $\heff$. The color of the symbols scale with the value of the 
semblance $\chi^2$, while the shape of the symbols change with the combination 
of parameters $(E,Ra)$ following the symbols already used in 
Fig.~\ref{fig:penet}. The errorbars correspond to one standard deviation about 
the mean values. The dashed horizontal lines show the nominal values 
for the geomagnetic field, while the blue shaded area correspond to one 
standard deviation in logarithmic scale \citep{Christensen10}. Given their poor 
Earth-likeness, the simulations with $E=3\times 10^{-4}$ have been excluded from 
these plots.}
 \label{fig:ratings}
\end{figure*}

For a more quantitative assessment, we now compare the morphology of the 
magnetic fields produced in the numerical models to the geomagnetic field at the 
CMB in terms of the four criteria introduced by \cite{Christensen10}. 
As shown in Fig.~\ref{fig:BrSmoothing}, the impact of the stable layer on the 
magnetic field morphology directly depends on the ratio $N_m/\Omega$ and hence 
on the distance of penetration $\dpen$ (Fig.~\ref{fig:penet}).
We now define a dynamical \emph{effective thickness} $\heff$ of the stable 
layer, which removes the distance of penetration of the convective eddies 
$\dpen$ from the actual static thickness $\hstrat$, such that

\begin{equation}
 \heff=\hstrat-\dpen=r_o-r_p\,.
 \label{eq:heff}
\end{equation}
We introduce this quantity to better capture the effective lengthscale that 
controls the magnetic field smoothing via the skin effect.
Figure~\ref{fig:ratings} shows the time-averages and the standard deviations of 
the four rating parameters AD/NAD, O/E, Z/NZ and FCF \citep{Christensen10} as a 
function of $\heff$. 
The series of numerical models with the highest 
Ekman number $E=3\times 10^{-4}$ and $Ra=3\times 10^6$ have been excluded from 
this plot since the fully convective simulation features a weakly-dipolar 
magnetic field and $\chi^2 > 8$. The relative axial dipole power AD/NAD 
(Fig.~\ref{fig:ratings}\textit{a}) is the criterion that shows the strongest 
dependence to the presence of a stable layer. The vast majority of 
the models with a thin or a vanishing stable layer (i.e. $\heff\simeq 0$~km)
indeed show AD/NAD values that lie within the $1\sigma$ tolerance level of
the nominal Earth's value. In contrast, the numerical models with $\heff > 
10$~km yield too dipolar magnetic field with AD/NAD ratios that grow well above 
the favoured value. In addition to the increase of AD/NAD, the stable 
stratification also makes the CMB magnetic field more antisymmetric with respect 
to the equator (Fig.~\ref{fig:ratings}\textit{b}) and more axisymmetric 
(Fig.~\ref{fig:ratings}\textit{c}), yielding O/E and Z/NZ ratios larger than the 
expected Earth's value. The flux concentration FCF shows a slightly 
different behaviour since weakly-stratified or fully convective models sometimes 
present ratios slightly larger than the nominal value, though they 
mostly lie within the $1\sigma$ tolerance range.

Overall the observed tendency is very similar for the four rating parameters: 
an increase of  $\heff$ goes along with a gradual smoothing of the CMB magnetic 
field which becomes more and more dipolar and axisymmetric. This analysis also 
demonstrates that $\heff$ is the key physical parameter that governs the 
Earth-likeness of the magnetic field independently of the variations of $E$, 
$Pm$ and $Ra$. The optimal numerical models which show the best 
agreement with 
the Earth CMB field in terms of $\chi^2$ values correspond to a vanishing 
effective thickness of the stable layer. This implies that to get a reasonable 
agreement with the geomagnetic field, the numerical models require either no 
stratified layer, or a penetration distance which is sufficient to span the 
entire static thickness of the layer. This yields the following upper bound for 
the thickness of the stable layer
\begin{equation}
 \hstrat \leq \dpen\,.
 \label{eq:maxhstrat}
\end{equation}
Using the scaling for the penetration distance (\ref{eq:fitpenet}), one gets

\begin{equation}
\hstrat \leq  3.2 
\left(\dfrac{\mathcal{N}}{\Omega}\bar{\ell}_s\right)^{-1}\,,
\label{eq:maxhstratnum}
\end{equation}
in dimensionless units. The above scaling relation could be further 
simplified by replacing $\bar{\ell}_s$ by the onset scaling obtained in 
Fig.~\ref{fig:forcebal}\textit{a}. Given the uncertainties when extrapolating 
numerical geodynamo models to Earth core conditions, we rather 
keep $\bar{\ell}_s$ for further discussion of
the geophysical implications of Eq.~(\ref{eq:maxhstratnum}).

\begin{figure}
 \centering
 \includegraphics[width=8.4cm]{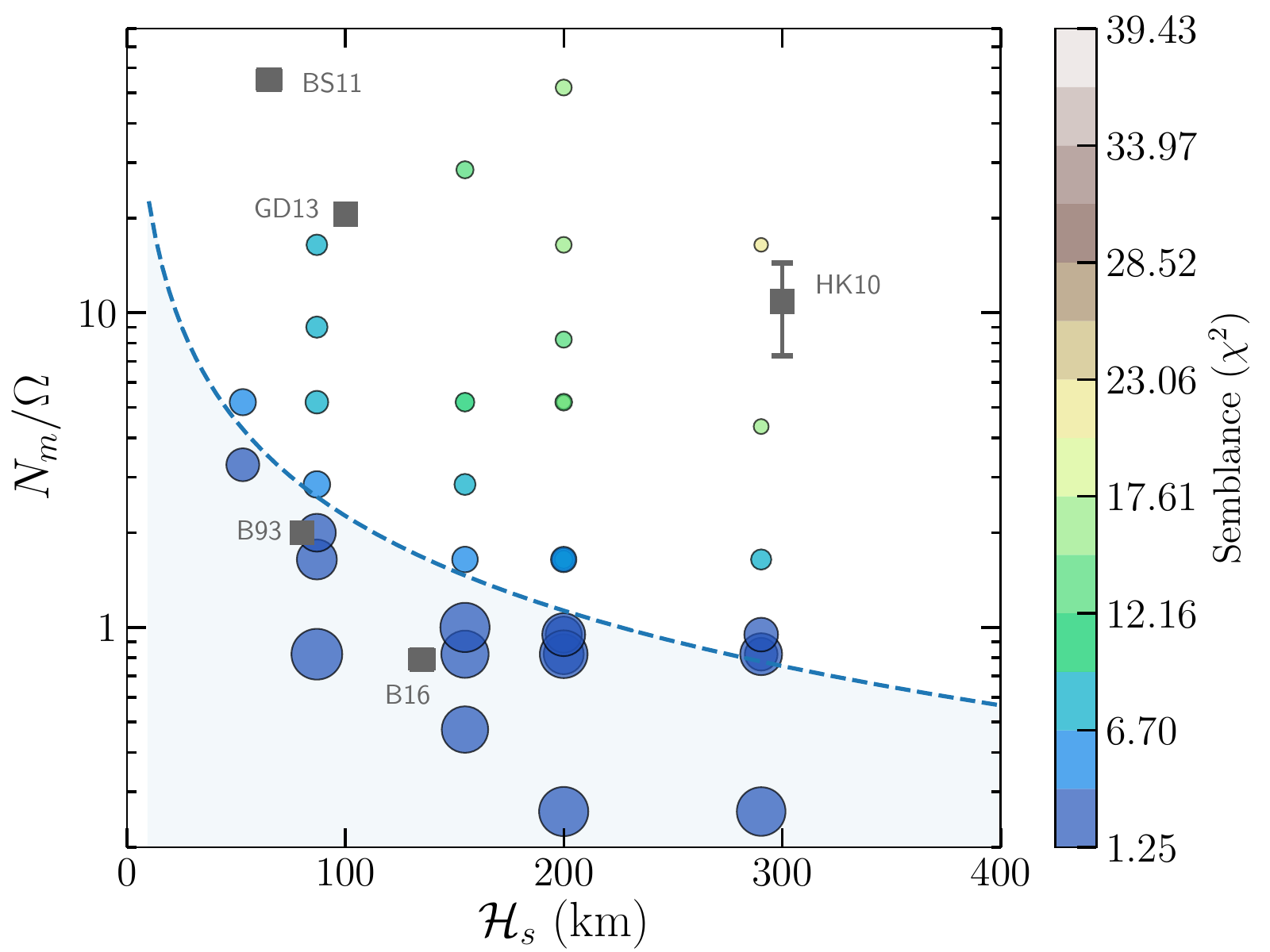}
 \caption{Morphological semblance between the numerical magnetic fields and the 
geomagnetic field at the CMB quantified by the measure of $\chi^2$ in the 
$(\hstrat,N_m/\Omega)$ parameter space for all the numerical 
simulations with fixed Ekman
and Rayleigh numbers ($E=3\times 10^{-5}$ and $Ra = 3\times 10^{8}$). The size 
of the symbols is inversely proportional to the value of $\chi^2$. 
The dashed blue line corresponds to the bound (\ref{eq:maxhstratnum}) 
derived using $\bar{\ell}_s=35$ (see Tab.~\ref{tab:results}) and assuming that 
$\mathcal{N}\simeq N_m$. The blue shaded region corresponds to the condition 
(\ref{eq:maxhstrat}). The different studies listed in Tab.~\ref{tab:strat} are 
marked by grey squares.}
  \label{fig:regimes}
\end{figure}

To further test the validity of this upper bound,
we focus on the 36 numerical simulations with $E=3\times 10^{-5}$ and 
$Ra=3\times 10^8$ for which the parameter space ($\hstrat$, $N_m/\Omega$) has 
been more densely sampled. 
Figure~\ref{fig:regimes} shows the morphological 
semblance $\chi^2$ in the $(\hstrat,N_m/\Omega)$ parameter space for this 
subset of simulations at fixed Ekman and Rayleigh numbers. For a 
practical determination of the upper bound given 
in Eq.~(\ref{eq:maxhstratnum}) and shown as a dashed line in 
Fig.~\ref{fig:regimes}, we use $\bar{\ell}_s=35$ (see Tab.~\ref{tab:results}) 
and make the assumption that $\mathcal{N}\simeq N_m$. The analysis of the 
distance of penetration (Fig.~\ref{fig:penet}\textit{a}) has already 
shown that this is a rather bold hypothesis that in practice yields some
dispersion of the data around the theoretical scaling (\ref{eq:dpen_takehiro}). 
This approximation is however mandatory for a comparison of the 
numerical models with the geophysical estimates. Indeed, while several studies 
suggest possible values of the maximum of the Brunt-V\"ais\"al\"a frequency 
$N_m$ for the Earth core (see Tab.~\ref{tab:strat}), $\mathcal{N}$ cannot be 
determined without the knowledge of $\dpen$, making its geophysical estimate 
rather uncertain. Despite this approximation, the scaling relation 
(\ref{eq:maxhstratnum})is found to correctly capture the transition 
between the numerical models with a good 
morphological agreement with the geomagnetic field (blue symbols with $\chi^2 < 
4$) from those which are non-compliant due their too dipolar 
structure.

\section{Geophysical implications}
\label{sec:geoph}

The condition (\ref{eq:maxhstrat}) puts a strong geophysical constraint on the 
acceptable degree of stratification.
For a comparison with 
the geophysical estimates of the physical properties of a stable layer at the 
top of the core coming from both seismic and magnetic studies, we report in 
Fig.~\ref{fig:regimes} the values of $\hstrat$ and $N_m/\Omega$ coming from the 
studies listed in Tab.~\ref{tab:strat}. 
Due to the magnetic field smoothing by skin effect, we fail to produce 
any Earth-like dynamo model with a stratification degree of $N_m/\Omega \geq 
10$ even for thicknesses as low as $\hstrat=50~$km.
Hence, a stable layer with $\hstrat \geq 100$~km and a stratification degree of 
$N_m/\Omega \simeq 10$ suggested by some seismic 
studies \citep[][]{Helffrich10,Tang15,Kaneshima18}
or of $N_m/\Omega > 20$ in models with a stable layer of compositional origin
\citep[][]{Buffet10,Gubbins13} seem hard to reconcile with our numerical 
geodynamo models. 
The condition (\ref{eq:maxhstrat}) can also be confronted to the estimates of 
outer core stratification that come from physical interpretation of the 
geomagnetic secular variation \citep{Brag93,Buffett16}.
% suggests that 
% the 60~yr period 
% observed in the geomagnetic secular variation would result from the excitation 
% of MAC waves in a stable layer underneath the CMB. Best-fitting models yield 
% $N_m/\Omega \simeq 1$ and $100<\hstrat <150$ \citep{Brag93,Buffett16}.
% 
% 
% 
% Stratification degree of $N_m/\Omega \simeq 10$
% A stratification degree of $N_m/\Omega \simeq 10$ obtained in some seismic 
% studies \citep[][]{Helffrich10,Tang15,Kaneshima18}
% or of $N_m/\Omega > 20$ in models with a stable layer of compositional origin
% \citep[][]{Buffet10,Gubbins13} seem hard to reconcile with the numerical 
% geodynamo models even for $\hstrat\leq 50$~km
% 
% in absence of an extra physical forcing in the stable layer 
% \cite[e.g.][]{Christensen18}.
% 
% In contrast, stable stratification of thermal origin with $N_m\sim \Omega$ and 
% $\hstrat \sim 100$~km allows the excitation of MAC waves that could 
% explain the 60~yr period observed in the geomagnetic secular variation 
% \citep{Brag93,Buffett14,Buffett16}.
In agreement with the previous findings by \cite{Olson17,Yan18} and 
\cite{Christensen18}, the numerical simulations with $E=3\times 10^{-5}$
yield an Earth-like magnetic field morphology when $\hstrat\sim 100$~km and 
$N_m\sim\Omega$.
% 
% Fig.~\ref{fig:regimes} shows that the combination of
% $\hstrat\sim 100$~km and $N_m\sim\Omega$ might be compatible with an Earth-like 
% magnetic field morphology, at least for simulations that have large Ekman 
% numbers. 
However, since the penetration distance directly depends on the 
horizontal lengthscale of the convective flow, the 
threshold obtained in Fig.~\ref{fig:regimes} using numerical 
simulations with $E=3\times 10^{-5}$ shall become more stringent at 
lower Ekman numbers when the convective flow lengthscale at $r_s$ is 
smaller.

\begin{figure}
 \centering
 \includegraphics[width=8.4cm]{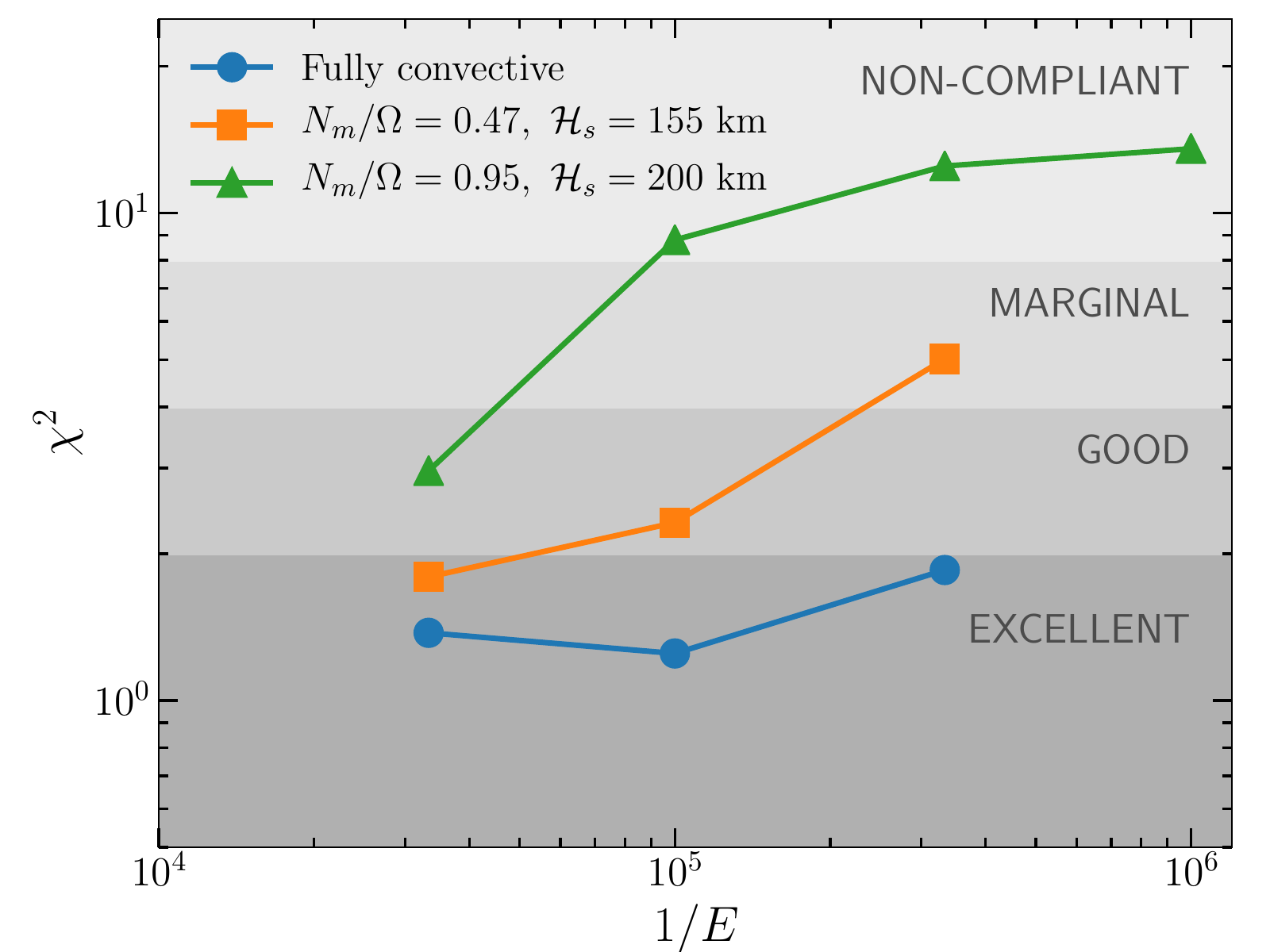}
 \caption{Compliance of field morphology quantified by its $\chi^2$ as 
a function of the Ekman number $E$ for three fully convective models (circles), 
three numerical models with $N_m/\Omega = 0.47$ and $\hstrat=155$~km 
(squares) and four numerical models with $N_m/\Omega=0.95$ and 
$\hstrat=200$~km (triangle)s. The grey shaded regions mark the boundaries of 
different levels of agreement with the Earth's magnetic field introduced by 
\cite{Christensen10}.}
 \label{fig:chi2_ek}
\end{figure}

To document this property, Fig.~\ref{fig:chi2_ek} shows the evolution of 
$\chi^2$ for three 
sets of numerical simulations with $N_m/\Omega\in[0,0.47,0.95]$ and 
$\hstrat\in[0,155,200]$~km for Ekman numbers decreasing  from $E=3\times 
10^{-5}$ to $E=10^{-6}$.
The numerical models which are fully convecting remain in excellent 
morphological agreement with the geomagnetic field (i.e. $\chi^2 < 2$) for the 
three Ekman numbers considered here. A closer inspection of
the four rating parameters however reveals a slow tendency to get more and more 
dipole-dominated magnetic fields when $E$ decreases. This increasing AD/NAD 
ratio is compensated by the evolution of FCF which is getting closer to
the expected Earth value at lower $E$. The numerical models with a stably 
stratified layer with a weak stratification $N_m/\Omega = 0.47$ or $N_m/\Omega = 
0.95$ show a stronger dependence to the Ekman number: while the $E=3\times 
10^{-5}$ cases still feature Earth-like magnetic fields, the compliance $\chi^2$ 
quickly degrades at lower $E$, yielding too dipolar and too axisymmetric 
magnetic fields incompatible with the geomagnetic observations. This is directly 
related to the decrease of the convective flow lengthscale which is found to 
follow $\bar{\ell}_s \sim E^{-1/3}$ atop the convective core 
(Fig.~\ref{fig:forcebal}\textit{a}). This goes along with smaller 
penetration distance $\dpen$ and hence larger $\heff$ which then yield an 
increased filtering of the CMB field by skin effect. 
For the lowest Ekman number 
considered here, we hence fail to produce an Earth-like magnetic field 
at a parameter combination $(\hstrat,N_m/\Omega)$ very close to the 
best-fitting models by \cite{Buffett16}. 

Dynamo models carry their own limitations and we can hence wonder whether 
there would be some leeway to viable (from a geomagnetic 
standpoint) stratification at Earth's core conditions. Here we envision three 
different scenarios to alleviate the severe 
limitation (\ref{eq:maxhstratnum}):

 \begin{description}

  \item[\emph{Larger distance of penetration}:]   A way to 
maintain $\heff=0$~km at a given value of $\hstrat$ would require an increase 
of the penetration distance. Based on the penetration distance of Alfv\'en 
waves, \cite{Takehiro15} for instance suggests that the hydrodynamical scaling 
(\ref{eq:dpen_takehiro}) should be replaced by
\[
 \dpen \sim \dfrac{Lu}{\bar{\ell}_s^2}, \quad 
Lu=\dfrac{2}{1+Pm}\left(\dfrac{\Lambda\,Pm}{E}\right)^{1/2}\,,
\]
when magnetic effects become important, $Lu$ being the Lundquist number 
\citep[e.g.][]{Schaeffer12}. At Earth's core conditions, this might 
yield much larger penetration distances than (\ref{eq:dpen_takehiro})  
\citep[see][]{Takehiro18}. Though a transition to the above scaling at a 
parameter range not covered in this study cannot be ruled out, our simulations 
do not show any correlation between the penetration distance $\dpen$ and 
the ratio $Lu/\bar{\ell}_s^2$. 
Furthermore, \cite{Takehiro15} specifically studied Alfv\'en wave 
penetration, which is rather different from the problem of penetrating 
convection in dynamo models.
In the latter, the hydromagnetic waves indeed exist at a significantly smaller 
level than the background magnetic field which is rather shaped by the slow 
convective motions \citep[e.g.][]{Hori15,Aubert18}. In this context, we do not 
anticipate that Alv\'en wave dynamics can have a significant impact on the 
attenuation properties of the background magnetic field.
% 
% 
% The simulations by \cite{Takehiro18} for which 
% the above scaling was reported to hold were conducted in the peculiar regime of 
% a forced uniform axial magnetic field with $\Lambda \leq 10^{-1}$ and 
% we might hence wonder about the applicability of Eq.~(\ref{eq:takalu}) to the 
% case of a dynamo-driven magnetic field with $\Lambda \sim 10$.
 
  \item[\emph{Larger convective flow lengthscale at $r_s$}:] The penetration 
distance directly depends on the horizontal lengthscale of the convective flow 
at the base of the stable layer $\mathcal{L}_s$. Given that the local 
convective supercriticality drops atop the convective core, 
the flow lengthscale at $r_s$ follows a local onset scaling of the 
form 
$\mathcal{L}_s \sim E^{1/3}$, or equivalently $\bar{\ell}_s \sim E^{-1/3}$.
At Earth's core conditions with $E=10^{-15}$ and $N_m\sim\Omega$, the 
penetration distance 
would be of the order $100$~m, would this onset scaling still hold.
Given the large diffusivities of the 3-D calculations, a transition to a 
magnetic control of $\bar{\ell}_s$ cannot be ruled out. The theoretical 
prediction by \cite{Davidson13} for a QG-MAC balance would then yield 
$\bar{\ell}_s \sim Ro^{-1/4}$ 
and hence $\dpen\sim 400$~km when using $Ro=Re E\sim 10^{-5}$ and 
$N_m\sim\Omega$.
However, while there is supporting evidence that the convective lengthscale in 
the bulk of the convective core departs from viscous control
\citep[see Fig.~\ref{fig:forcebal}\textit{a} and][]{Aubert17,Schwaiger19}, our 
simulations do not suggest that the 
interface flow at $r_s$ should follow the same scaling.

\item[\emph{Additional physical forcings in the stable layer}:]
The last avenue to alleviate the criterion 
(\ref{eq:maxhstratnum}) relies on additional forcings to drive flows in 
the stably-stratified layer. In contrast to the assumptions made in this 
study, the CMB heat flow is expected to be strongly 
heterogeneous and hence drive flows by thermal winds. Using dynamo models with 
a stable layer with $N_m/\Omega \leq 4$ and an heterogeneous heat flux pattern, 
\cite{Christensen18} has derived a scaling relation for the flow driven by the
CMB thermal heterogeneities. At Earth's core conditions, this flow is expected 
to be very shallow limited to the first few hundred meters below the CMB and 
might hence have a moderate impact on the magnetic field morphology, 
would the extrapolation from geodynamo simulations to Earth condition 
holds.
Because of the strong core-mantle heat flux 
heterogeneities, the stratification might not be global but rather confined 
to localised regions as suggested by the hydrodynamical numerical simulations 
by \cite{Mound19}. Regional stratification could however yield a heterogeneous
magnetic field at the CMB with a weaker field with a smoother morphology in the 
stratified area. The viability of this scenario remains hence to be assessed 
by means of global geodynamo models.
Other physical forcings not accounted for in our models, such as 
double-diffusive effects, could possibly impact the dynamics of the 
outer layer. A promising physical configuration arises when thermal
stratification is stable while compositional stratification is unstable, a 
configuration akin to fingering convection that develops in the ocean when warm 
and salty water lies above cold and fresh water \citep[e.g.][]{Radko13}.
Numerical models by \cite{Manglik10} and \cite{Takahashi19}, carried 
out in the context of modelling 
Mercury's dynamo, indicate that fingering convection enhances the
convective penetration in the thermally-stratified layer when $N_m\sim \Omega$
\citep[see also][]{Monville19,Silva19,Bouffard19}.
% 
% the upwellings that 
% penetrate the external layer are enhanced
% 
% For moderate stratification of thermal origin with $N_m\sim 
% \Omega$, 
% 
% \citep{Manglik10,Bouffard19
% 
% and possibly modifiy the criterion (\ref{eq:maxhstrat}).
 \end{description}

% Optimum suggest $\heff=0$~km
% Dynamo models privilege 0~km.
% 
% $\bar{\ell}_s \sim E^{1/3}$ yield $\dpen=2$~km 
% $\bar{\ell}_s \sim Ro^{-1/4}$ yield $\dpen\sim 200$~km but there is no 
% indication that the convective lengthscale at the base of the stable layer 
% follows this scaling relation.

% \begin{figure}
%  \centering
%  \includegraphics[width=8.4cm]{compEarth}
%  \caption{Normalised time-averaged magnetic field spectra at the CMB as a 
% function of the spherical harmonic degree $\ell$. The lines with square and 
% triangle markers correspond to numerical models with $E=3\times 10^{-5}$, 
% $Ra=3\times 10^8$, $Pm=2.5$ and $r_s=1.5$ (i.e. $\hstrat = 87$~km). The thick 
% line with circular markers correspond to the normalised spectra 
% from the COV-OBS model time-averaged over the period 1900-2010. The standard 
% deviation denoted by a shaded region incorporate both the temporal variations 
% and the model uncertainties.}
%  \label{fig:spec_Earth}
% \end{figure}
% 
% When discussing the final spectra figure, one can justify the normalisation 
% following \cite{Christensen18} who claim there a leeway to rescale when using 
% power-based spectra.
% Take-away on this figure: FC and $N_m/\Omega = 0.82$ are indiscernible. Beyond 
% $N_m/\Omega \simeq 5$ the decay is too strong. Quadrupole/Dipole ratio and 
% Octupole/Dipole ratio are key to constrain the presence of a stably-stratified 
% layer \citep{Olson17,Yan18}

\section{Conclusion}
\label{sec:concl}

In this study, we have examined the physical effect of a stably-stratified 
layer underneath the core-mantle boundary by means of 3-D global geodynamo 
simulations in spherical geometry. We have introduced a parametrised 
temperature background to independently vary the thickness 
$\hstrat$ and the degree of stratification of the stable layer, quantified here 
by the ratio of the maximum Brunt-V\"ais\"al\"a frequency over the rotation 
rate $N_m/\Omega$.
We have conducted a systematic survey by varying $\hstrat$ from $0$ to $290$~km 
and 
$N_m/\Omega$ from $0$ to more than $50$ for several combinations of Ekman and 
Rayleigh numbers. This parameter range encompasses the possible values of 
the physical properties of a stable layer underneath the CMB that come either 
from seismic or from geomagnetic studies (see Tab.~\ref{tab:strat}).
This work complements previous analyses that were either limited to moderate 
stratification degree $N_m/\Omega< 5$ \citep{Olson17,Yan18,Christensen18} or to 
moderate control parameters like large Ekman numbers \citep[$E=3\times 
10^{-4}$, ][]{Nakagawa11} or dynamo action close to onset \citep{Nakagawa15}.

We have first studied the penetration of the convective motions in the 
stably-stratified layer. When using the radial profile of the buoyancy power to 
define the penetration distance $\dpen$, we have shown that $\dpen \sim 
(\mathcal{N}\bar{\ell}_s/\Omega)^{-1}$ where $\mathcal{N}$ incorporates the 
local variation of the Brunt-V\"ais\"al\"a frequency and $\bar{\ell}_s$ relates 
to the typical size of the convective eddies $\mathcal{L}_s$ at the top of the 
convective core 
via $\bar{\ell}_s=\pi r_s/\mathcal{L}_s$. This scaling is in perfect agreement 
with the theoretical prediction by \cite{Takehiro01} which has been derived in 
absence of magnetic effects. Because of the drop 
of the convective supercriticality at the top of the convective core,
the convective lengthscale at the transition radius $r_s$ has been found to 
follow an onset scaling, i.e. $\mathcal{L}_s \sim E^{1/3}$.
Our results hence indicate that the magnetic field 
has little influence on the penetration distance, in contrast with 
the theoretical expectations by \cite{Takehiro15}.
To explain this somewhat surprising result, we note that when the magnetic field 
is self-sustained -as opposed to the imposed field considered by 
\cite{Takehiro18}-, hydromagnetic waves have a much weaker amplitude
than the background magnetic field which is rather shaped by the slow 
convective motions \citep[e.g.][]{Hori15}. We hence anticipate that the 
dynamics of the Alv\'en waves at $r_s$ have little impact on the distance of 
penetration of the convective features.

% 
% dynamo models with
% a strong field with $\Lambda \sim 10$ still follow this hydrodynamical scaling.
% In contrast to the theoretical work by \cite{Takehiro15}, the Lorentz force
% plays a negligable role here in controlling the 
% penetration distance $\dpen$.

Stable stratification has a strong impact on the magnetic field morphology at 
the CMB. Because of vanishing convective flows in the stable 
layer, the small-scale features of the magnetic field are smoothed out by
skin effect 
\cite[e.g.][]{Christensen06,Gubbins07}.
% The stably stratified layer acts as a low-pass filter on the magnetic 
% field, smoothing out the small-scale features.
Using the rating parameters defined by \cite{Christensen10} to assess the Earth 
likeness of the numerical models fields, we have shown that the physically 
relevant lengthscale is the \emph{effective thickness} of the stable layer 
$\heff$, which results from the difference between the actual static thickness 
$\hstrat$ and the penetration distance $\dpen$.  Only models with a 
vanishing $\heff$ yield a good agreement with the Earth CMB field. This 
implies that Earth-like dynamo models either harbour a 
fully-convecting core or have a penetration distance which is sufficient to 
cross the entire stable layer. The combination of the scaling obtained for the 
penetration distance $\dpen$ and the condition $\heff=0$~km yields the 
following upper bound for the thickness of the stable layer underneath the CMB

\[
 \hstrat \leq \left(\dfrac{N_m}{\Omega}\right)^{-1} \mathcal{L}_s\,.
\]
This condition puts severe
limitations on the acceptable degree of stratification. Large degrees of 
$N_m/\Omega \sim 10$ suggested by several seismic studies 
\citep[e.g.][]{Helffrich10} yield magnetic field morphology that are 
incompatible with the geomagnetic field observations at the CMB even for a layer 
as small as $\hstrat=50$~km. In agreement with previous findings by 
\cite{Olson17} and \cite{Christensen18}, we have shown that geodynamo models 
with a smaller stratification $N_m\sim\Omega$ and $\hstrat\sim 100$~km sustain 
a magnetic field morphology that is compatible with the geomagnetic 
observations, as long as the Ekman number is large enough, i.e. $E \geq 
3\times 10^{-5}$. Since the convective lengthscale at the top of the convective 
core decreases with the Ekman number, following the 
onset scaling $\mathcal{L}_s \sim E^{1/3}$, 
the penetration distance decreases and the Earth-likeness of the numerical 
models fields degrades. At Earth's core 
conditions with $E=10^{-15}$ and $N_m\sim \Omega$, the penetration distance 
could be reduced to hundreds of meter, yielding a strong magnetic skin effect 
incompatible with geomagnetic observations.

Consequently, our suite of numerical models, given the type and magnitude 
of physical processes governing the dynamics of the stably stratified layer 
that they incorporate, favour the absence of stable stratification atop Earth's 
core.
% 
% hence favour the 
% absence of stable stratification}
% 
% Our numerical geodynamo models hence favour no stable stratification atop
% the core \citep[e.g.][]{Irving18}, unless unaccounted physical forcings
% modify the dynamics of the stable layer.
% 
% double-diffusive effects 
% modify the dynamics of the stable layer. Indeed, while we have assumed a pure 
% thermal stratification in this study,  the layer underneath the CMB might 
% rather operate in the \emph{fingering convection} regime in which
% compositional effects are destabilising while thermal effects are stabilising
% \citep{Manglik10,Takahashi19}. The influence of a magnetic field on rotating 
% fingering convection remains largely unexplored 
% \citep[e.g.][]{Monville19,Silva19,Bouffard19} and requires a dedicated 
% study.

\begin{acknowledgments}
We would like to thank Maylis Landeau and Vincent Lesur for fruitful 
discussions during the elaboration of this manuscript.
We acknowledge support from the Fondation Simone and Cino Del Duca of Institut 
de France (JA, 2017 Research Grant) and from CNRS Programme National de 
Plan\'etologie (TG, 2017 PNP Grant). Numerical computations have been carried 
out on the S-CAPAD platform at IPGP and using HPC resources from GENCI (Grants 
A0040402122, A0070410095 and 2019gch0411).  All the figures 
have been generated using \texttt{matplotlib} \citep{Hunter07} and 
\texttt{paraview} (\url{https://www.paraview.org}).
\end{acknowledgments}

\bibliographystyle{gji}
%\bibliography{biblio}

\appendix

% {\onecolumn
% \begin{small}
% \begin{planotable}{rrrrrrrrrrrrrrrr}
% \tablecaption{This is the caption}
% \tablehead{$E$ & $Ra$ & $Pm$ & $\hstrat$ & $N_m/\Omega$ & $Rm$ & $\Lambda$ & 
% $\bar{\ell}_s$ 
% &  $\dpen$ & AD/NAD & O/E & Z/NZ & FCF & $\chi^2$ & $(N_r,\ell_{\text{max}})$ & 
% $t_\text{run}$}
% \startdata
% \multicolumn{16}{c}{he he he} \nl
% \input{plano}
% \end{planotable}
% \end{small}}

{\onecolumn
%{\footnotesize
\begin{center}
 \input{tab_results.tex}
\end{center}
%}
}

\bsp 

\label{lastpage}

\end{document}

%% file: tab_results.tex
\begin{longtable}{rrrrrrrrrrrrrrrl}
\caption{Table of model parameters and results. The distances
$\hstrat$ and $\dpen$ are expressed in kilometers. The total run
time $t_\text{run}$ is given in magnetic diffusion time.
All simulations have assumed $Pr=1$. The numerical simulations with an asterisk 
in the last column have been computed with the \texttt{PARODY-JA} code.}
\label{tab:results} \\

\toprule
$Pm$ & $\hstrat$ & $N_m/\Omega$ & $Rm$ & $\Lambda$ & $\bar{\ell}_s$ &  $\dpen$ & AD/NAD & O/E & Z/NZ & FCF & $\chi^2$ & $N_r$ & $\ell_{\text{max}}$ & $\alpha_{\text{map}}$  & $t_\text{run}$ \\
\midrule
\endfirsthead

\midrule
\multicolumn{16}{c}{\tablename\ \thetable\ -- Continued from previous page} \\
\midrule
\endhead

\midrule
\multicolumn{16}{c}{\tablename\ \thetable\ -- Continued on next page} \\
\midrule
\endfoot

\bottomrule
\endlastfoot
\multicolumn{8}{r}{$E=10^{-3}$} & \multicolumn{8}{l}{$Ra=3\times 10^{5}$} \\
$15.00$ & $200$ & $0.94$ & $366$ & $7.5$ & $7$ & $200$ & $0.03$ & $1.34$ & 
$0.12$ & $6.25$ & $39.4$ & $49$ & $85$ & - & $5.40$ \\
\multicolumn{8}{r}{$E=3\times 10^{-4}$} & \multicolumn{8}{l}{$Ra=3\times 10^{6}$} \\
$5.00$ & $0$  & - & $252$ & $15.0$ &- & - & $0.49$ & $2.14$ & $0.31$ & $4.89$ & $8.6$ & $65$ & $106$ & $0.86$ & $1.12$ \\
$5.00$ & $155$ & $4.35$ & $225$ & $11.2$ & $15$ & $131$ & $1.94$ & $3.53$ & $0.50$ & $2.07$ & $5.6$ & $81$ & $106$ & - & $1.60$ \\
$5.00$ & $200$ & $0.82$ & $248$ & $13.6$ & $13$ & $200$ & $0.61$ & $2.42$ & $0.33$ & $4.43$ & $7.5$ & $81$ & $106$ & - & $1.15$ \\
$5.00$ & $200$ & $0.95$ & $246$ & $13.9$ & $13$ & $200$ & $0.68$ & $2.18$ & $0.32$ & $4.08$ & $6.3$ & $65$ & $106$ & - & $1.11$ \\
$5.00$ & $200$ & $4.35$ & $214$ & $12.4$ & $14$ & $161$ & $2.56$ & $3.41$ & $0.48$ & $1.78$ & $5.6$ & $81$ & $106$ & - & $2.14$ \\
$5.00$ & $290$ & $0.82$ & $240$ & $13.9$ & $12$ & $290$ & $0.74$ & $2.57$ & $0.31$ & $3.92$ & $6.3$ & $81$ & $106$ & - & $1.50$ \\
$5.00$ & $290$ & $1.37$ & $229$ & $12.9$ & $13$ & $270$ & $1.11$ & $3.37$ & $0.31$ & $3.24$ & $5.7$ & $81$ & $106$ & - & $2.03$ \\
$5.00$ & $290$ & $4.35$ & $202$ & $12.7$ & $13$ & $179$ & $3.01$ & $3.88$ & $0.52$ & $1.62$ & $6.9$ & $97$ & $106$ & - & $1.36$ \\
$5.00$ & $290$ & $7.35$ & $196$ & $12.7$ & $13$ & $129$ & $3.30$ & $4.33$ & $0.55$ & $1.54$ & $8.0$ & $129$ & $106$ & - & $1.31$ \\
$5.00$ & $290$ & $13.75$ & $189$ & $12.7$ & $13$ & $92$ & $4.02$ & $3.56$ & $0.56$ & $1.39$ & $7.8$ & $145$ & $106$ & - & $1.71$ \\
$5.00$ & $290$ & $23.24$ & $186$ & $12.6$ & $13$ & $69$ & $4.28$ & $4.22$ & $0.63$ & $1.41$ & $9.4$ & $145$ & $106$ & - & $1.12$ \\
$5.00$ & $290$ & $43.47$ & $183$ & $12.1$ & $13$ & $48$ & $4.45$ & $4.03$ & 
$0.94$ & $1.44$ & $10.9$ & $193$ & $106$ & - & $1.13$ \\
\multicolumn{8}{r}{$E=10^{-4}$} & \multicolumn{8}{l}{$Ra=4\times 10^{7}$} \\
$3.50$ & $200$ & $0.95$ & $407$ & $19.4$ & $20$ & $191$ & $1.19$ & $2.02$ & 
$0.38$ & $2.63$ & $3.1$ & $81$ & $106$ & - & $1.30$ \\
\multicolumn{8}{r}{$E=3\times 10^{-5}$} & \multicolumn{8}{l}{$Ra=10^{8}$} \\
$2.50$ & $0$  & - & $302$ & $17.9$ &- & - & $1.31$ & $1.53$ & $0.19$ & $3.65$ & $3.0$ & $81$ & $106$ & $0.91$ & $1.12$ \\
$2.50$ & $200$ & $0.47$ & $288$ & $16.1$ & $25$ & $193$ & $1.86$ & $1.51$ & $0.17$ & $2.80$ & $1.8$ & $81$ & $106$ & $0.91$ & $1.05$ \\
$2.50$ & $200$ & $0.82$ & $292$ & $12.4$ & $26$ & $191$ & $3.53$ & $2.00$ & $0.29$ & $1.58$ & $3.3$ & $81$ & $106$ & $0.91$ & $1.08$ \\
$2.50$ & $200$ & $1.64$ & $282$ & $11.5$ & $27$ & $161$ & $6.10$ & $2.73$ & $0.42$ & $1.20$ & $8.0$ & $81$ & $106$ & $0.91$ & $1.04$ \\
\multicolumn{8}{r}{$E=3\times 10^{-5}$} & \multicolumn{8}{l}{$Ra=3\times 10^{8}$} \\
$1.00$ & $0$  & - & $234$ & $7.6$ &- & - & $2.59$ & $1.59$ & $0.24$ & $1.87$ & $1.7$ & $81$ & $128$ & - & $1.00$ \\
$1.00$ & $200$ & $1.64$ & $213$ & $6.9$ & $34$ & $188$ & $9.29$ & $2.50$ & $0.51$ & $1.04$ & $11.4$ & $81$ & $128$ & - & $1.29$ \\
$1.00$ & $200$ & $5.20$ & $205$ & $6.3$ & $35$ & $87$ & $16.35$ & $2.77$ & $0.74$ & $0.86$ & $18.8$ & $81$ & $128$ & - & $1.03$ \\
$2.50$ & $0$  & - & $555$ & $23.2$ &- & - & $1.54$ & $1.58$ & $0.28$ & $2.19$ & $1.4$ & $81$ & $128$ & $0.91$ & $1.04$ \\
$2.50$ & $53$ & $3.29$ & $550$ & $18.1$ & $36$ & $47$ & $3.08$ & $1.98$ & $0.43$ & $1.56$ & $3.6$ & $81$ & $128$ & $0.91$ & $1.09$ \\
$2.50$ & $53$ & $5.20$ & $543$ & $17.1$ & $41$ & $45$ & $3.85$ & $2.35$ & $0.53$ & $1.31$ & $5.6$ & $145$ & $128$ & $0.97$ & $1.13$ \\
$2.50$ & $87$ & $0.82$ & $546$ & $22.8$ & $31$ & $83$ & $1.70$ & $1.59$ & $0.30$ & $2.12$ & $1.5$ & $81$ & $128$ & $0.91$ & $1.03$ \\
$2.50$ & $87$ & $1.64$ & $548$ & $19.9$ & $34$ & $81$ & $2.47$ & $1.90$ & $0.34$ & $1.76$ & $2.4$ & $81$ & $128$ & $0.91$ & $1.18$ \\
$2.50$ & $87$ & $2.85$ & $535$ & $18.2$ & $38$ & $79$ & $3.91$ & $2.25$ & $0.53$ & $1.40$ & $5.5$ & $81$ & $128$ & - & $0.82$ \\
$2.50$ & $87$ & $5.20$ & $533$ & $17.4$ & $40$ & $62$ & $5.35$ & $2.40$ & $0.53$ & $1.12$ & $7.5$ & $97$ & $128$ & $0.93$ & $1.38$ \\
$2.50$ & $87$ & $9.00$ & $522$ & $17.6$ & $39$ & $39$ & $6.82$ & $2.21$ & $0.46$ & $1.02$ & $8.5$ & $161$ & $128$ & $0.97$ & $1.01$ \\
$2.50$ & $87$ & $16.43$ & $517$ & $18.1$ & $37$ & $27$ & $7.70$ & $2.17$ & $0.40$ & $0.97$ & $9.0$ & $161$ & $128$ & $0.97$ & $1.10$ \\
$2.50$ & $155$ & $0.47$ & $541$ & $23.3$ & $29$ & $155$ & $1.63$ & $1.73$ & $0.30$ & $2.25$ & $1.8$ & $81$ & $128$ & $0.91$ & $1.10$ \\
$2.50$ & $155$ & $0.82$ & $535$ & $22.4$ & $31$ & $149$ & $2.07$ & $1.71$ & $0.29$ & $2.02$ & $1.7$ & $81$ & $128$ & $0.91$ & $1.05$ \\
$2.50$ & $155$ & $1.64$ & $527$ & $18.9$ & $34$ & $144$ & $4.16$ & $2.35$ & $0.53$ & $1.39$ & $5.9$ & $81$ & $128$ & - & $1.17$ \\
$2.50$ & $155$ & $2.85$ & $516$ & $18.5$ & $36$ & $111$ & $6.00$ & $2.49$ & $0.60$ & $1.14$ & $8.7$ & $81$ & $128$ & $0.91$ & $1.16$ \\
$2.50$ & $155$ & $5.20$ & $503$ & $18.7$ & $36$ & $75$ & $8.12$ & $2.56$ & $0.58$ & $1.01$ & $10.9$ & $145$ & $128$ & $0.96$ & $1.00$ \\
$2.50$ & $155$ & $28.46$ & $487$ & $18.5$ & $35$ & $25$ & $10.95$ & $2.22$ & $0.52$ & $0.85$ & $13.0$ & $145$ & $133$ & $0.97$ & $1.15$ \\
$2.50$ & $200$ & $0.26$ & $536$ & $24.2$ & $30$ & $200$ & $1.62$ & $1.66$ & $0.27$ & $2.31$ & $1.6$ & $81$ & $128$ & $0.91$ & $1.07$ \\
$2.50$ & $200$ & $0.82$ & $526$ & $21.5$ & $31$ & $193$ & $2.41$ & $1.89$ & $0.34$ & $1.90$ & $2.4$ & $81$ & $128$ & $0.91$ & $1.06$ \\
$2.50$ & $200$ & $0.95$ & $528$ & $20.2$ & $31$ & $192$ & $2.86$ & $1.96$ & $0.36$ & $1.73$ & $3.0$ & $81$ & $133$ & $0.91$ & $1.40$ \\
$2.50$ & $200$ & $1.64$ & $519$ & $18.7$ & $33$ & $182$ & $4.48$ & $2.90$ & $0.59$ & $1.33$ & $7.4$ & $81$ & $128$ & - & $1.51$ \\
$2.50$ & $200$ & $5.20$ & $489$ & $18.4$ & $35$ & $88$ & $9.87$ & $2.89$ & $0.65$ & $0.93$ & $13.6$ & $145$ & $170$ & $0.97$ & $1.03$ \\
$2.50$ & $200$ & $8.22$ & $485$ & $18.4$ & $36$ & $64$ & $11.80$ & $2.60$ & $0.59$ & $0.86$ & $14.6$ & $145$ & $128$ & - & $1.10$ \\
$2.50$ & $200$ & $16.43$ & $477$ & $17.9$ & $36$ & $40$ & $12.28$ & $2.68$ & $0.57$ & $0.79$ & $15.3$ & $145$ & $128$ & $0.97$ & $1.19$ \\
$2.50$ & $200$ & $51.96$ & $465$ & $17.5$ & $35$ & $19$ & $12.97$ & $2.16$ & $0.57$ & $0.79$ & $15.0$ & $257$ & $170$ & $0.98$ & $1.03$ \\
$2.50$ & $290$ & $0.26$ & $524$ & $24.3$ & $28$ & $290$ & $1.61$ & $1.60$ & $0.28$ & $2.38$ & $1.6$ & $81$ & $128$ & $0.91$ & $1.07$ \\
$2.50$ & $290$ & $0.82$ & $515$ & $19.6$ & $30$ & $281$ & $3.10$ & $2.14$ & $0.39$ & $1.67$ & $3.6$ & $81$ & $128$ & $0.91$ & $1.44$ \\
$2.50$ & $290$ & $4.35$ & $462$ & $17.9$ & $32$ & $119$ & $13.94$ & $2.86$ & $0.66$ & $0.79$ & $17.2$ & $81$ & $128$ & $0.91$ & $1.08$ \\
$2.50$ & $290$ & $16.43$ & $447$ & $16.5$ & $37$ & $47$ & $17.04$ & $2.98$ & 
$0.76$ & $0.70$ & $20.4$ & $145$ & $128$ & $0.97$ & $1.10$ \\
$4.33$ & $0$  & - & $935$ & $46.2$ &- & - & $1.23$ & $1.49$ & $0.27$ & $2.40$ & 
$1.5$ & $160$ & $133$ & - & $0.48^\star$ \\
$4.33$ & $87$ & $2.00$ & $908$ & $38.0$ & - & $79$ & $2.36$ & $1.90$ & 
$0.40$ & $1.83$ & $2.7$ & $160$ & $133$ & - & $0.55^\star$ \\
$4.33$ & $155$ & $1.00$ & $891$ & $42.7$ & - & $147$ & $1.85$ & $1.76$ & $0.29$ 
& $2.04$ & $1.6$ & $160$ & $133$ & - & $0.54^\star$ \\
$4.33$ & $200$ & $0.82$ & $888$ & $43.1$ & - & $192$ & $1.68$ & $1.71$ & $0.27$ 
& $2.35$ & $1.7$ & $160$ & $133$ & - & $0.71^\star$ \\
$4.33$ & $200$ & $0.95$ & $879$ & $41.4$ & - & $192$ & $2.08$ & $1.86$ & $0.32$ 
& $1.98$ & $2.1$ & $160$ & $133$ & - & $0.73^\star$ \\
$4.33$ & $200$ & $1.64$ & $867$ & $37.0$ & - & $174$ & $3.97$ & $2.61$ & $0.51$ 
& $1.36$ & $6.0$ & $160$ & $133$ & - & $0.68^\star$ \\
$4.33$ & $290$ & $0.82$ & $863$ & $40.2$ & - & $278$ & $2.23$ & $2.01$ & $0.30$ 
& $1.90$ & $2.2$ & $160$ & $133$ & - & $0.74^\star$ \\
$4.33$ & $290$ & $0.95$ & $858$ & $38.0$ & - & $278$ & $2.64$ & $2.37$ & $0.37$ 
& $1.76$ & $3.4$ & $160$ & $133$ & - & $0.67^\star$ \\
$4.33$ & $290$ & $1.64$ & $820$ & $36.9$ & - & $224$ & $6.99$ & $2.64$ & $0.48$ 
& $1.02$ & $9.4$ & $160$ & $133$ & - & $0.64^\star$ \\ 
\multicolumn{8}{r}{$E=3\times 10^{-5}$} & \multicolumn{8}{l}{$Ra=10^{9}$} \\
$1.44$ & $0$  & - & $617$ & $17.3$ &- & - & $1.88$ & $2.34$ & $0.58$ & $1.74$ & $3.9$ & $97$ & $170$ & $0.93$ & $1.09$ \\
$1.44$ & $155$ & $0.87$ & $596$ & $17.2$ & $33$ & $150$ & $2.33$ & $2.31$ & $0.64$ & $1.71$ & $4.6$ & $97$ & $170$ & $0.93$ & $1.07$ \\
$1.44$ & $155$ & $1.73$ & $583$ & $16.2$ & $36$ & $148$ & $3.79$ & $2.73$ & $0.68$ & $1.49$ & $6.9$ & $97$ & $170$ & - & $1.47$ \\
$1.44$ & $200$ & $0.87$ & $587$ & $17.4$ & $33$ & $194$ & $2.70$ & $2.35$ & $0.60$ & $1.67$ & $4.8$ & $97$ & $170$ & $0.93$ & $1.08$ \\
$1.44$ & $200$ & $1.73$ & $578$ & $15.5$ & $36$ & $191$ & $4.57$ & $3.15$ & $0.78$ & $1.33$ & $8.9$ & $97$ & $170$ & $0.93$ & $1.22$ \\
$1.44$ & $290$ & $1.73$ & $561$ & $14.9$ & $34$ & $275$ & $5.67$ & $3.93$ & $0.81$ & $1.18$ & $11.5$ & $97$ & $170$ & $0.97$ & $1.17$ \\
\multicolumn{8}{r}{$E=10^{-5}$} & \multicolumn{8}{l}{$Ra=2\times 10^{9}$} \\
$1.20$ & $0$  & - & $442$ & $15.7$ &- & - & $2.48$ & $1.52$ & $0.19$ & $1.85$ & $1.2$ & $129$ & $192$ & $0.96$ & $1.05$ \\
$1.20$ & $155$ & $0.47$ & $429$ & $15.3$ & $43$ & $151$ & $3.57$ & $1.53$ & $0.20$ & $1.57$ & $2.3$ & $129$ & $192$ & $0.96$ & $1.08$ \\
$1.20$ & $200$ & $0.95$ & $415$ & $13.8$ & $41$ & $189$ & $8.44$ & $2.00$ & $0.33$ & $1.07$ & $8.8$ & $129$ & $192$ & $0.96$ & $1.01$ \\
\multicolumn{8}{r}{$E=3\times 10^{-6}$} & \multicolumn{8}{l}{$Ra=10^{10}$} \\
$0.80$ & $0$  & - & $387$ & $13.1$ &- & - & $3.29$ & $1.40$ & $0.12$ & $1.71$ & $1.9$ & $161$ & $256$ & $0.97$ & $0.67$ \\
$0.80$ & $155$ & $0.47$ & $375$ & $12.0$ & $55$ & $150$ & $6.30$ & $1.31$ & $0.17$ & $1.22$ & $5.0$ & $161$ & $256$ & $0.97$ & $0.66$ \\
$0.80$ & $200$ & $0.95$ & $388$ & $10.0$ & $59$ & $152$ & $10.43$ & $3.11$ & $0.31$ & $0.92$ & $12.5$ & $193$ & $256$ & $0.98$ & $0.60$ \\
\multicolumn{8}{r}{$E=3\times 10^{-6}$} & \multicolumn{8}{l}{$Ra=3\times 10^{10}$} \\
$0.80$ & $155$ & $1.64$ & $651$ & $17.6$ & $65$ & $110$ & $10.78$ & $2.13$ & 
$0.42$ & $0.85$ & $12.2$ & $193$ & $288$ & $0.98$ & $0.51$ \\
%   \multicolumn{8}{r}{$E=10^{-6}$} & \multicolumn{8}{l}{$Ra=9\times 10^{10}$} \\
% $0.50$ & $200$ & $0.95$ & $478$ & $8.7$ & $86$ & $140$ & $4.44$ & $5.09$ & 
% $0.51$ & $1.02$ & $10.6$ & $321$ & $426$ & $0.99$ & $0.13$ \\
 \multicolumn{8}{r}{$E=10^{-6}$} & \multicolumn{8}{l}{$Ra=9\times 10^{10}$} \\
$0.50$ & $200$ & $0.95$ & $461$ & $9.1$ & $74$ & $152$ & $5.09$ & $6.94$ & 
$0.50$ & $1.00$ & $13.6$ & $321$ & $426$ & $0.99$ & $0.23$ \\
%   \multicolumn{8}{r}{$E=3\times 10^{-7}$} & \multicolumn{8}{l}{$Ra=9\times 
% 10^{11}$} \\
% $0.25$ & $200$ & $0.95$ & $483$ & $6.3$ & $129$ & $126$ & $4.84$ & $4.41$ & 
% $0.71$ & $0.72$ & $12.4$ & $577$ & $597$ & $0.99$ & $0.03$ \\
\end{longtable}